\documentclass[twocolumn]{aastex62}

\usepackage{graphicx}
\usepackage{longtable}
\usepackage{amsmath}
\usepackage{booktabs}
\usepackage{multirow}
\usepackage{float}

\graphicspath{{./}{figures/}}

\submitjournal{ApJ}

\shorttitle{Hidden BLR in Bulgeless Galaxy}
\shortauthors{Bohn et al.}

\begin{document}

\title{The Discovery of a Hidden Broad-line AGN in a Bulgeless Galaxy: Keck NIR Spectroscopic Observations of SDSS J085153.64+392611.76}

\email{tbohn002@ucr.edu}

\author[0000-0002-4375-254X]{Thomas Bohn}
\affil{University of California, Riverside, Department of Physics \& Astronomy \\
900 University Ave., Riverside, CA 92521, USA}

\author[0000-0003-4693-6157]{Gabriela Canalizo}
\affiliation{University of California, Riverside, Department of Physics \& Astronomy \\
900 University Ave., Riverside, CA 92521, USA}

\author[0000-0003-2277-2354]{Shobita Satyapal}
\affiliation{George Mason University, Department of Physics \& Astronomy \\
MS3F3, 4400 University Drive, Fairfax, VA 22030, USA}

\author[0000-0001-8640-8522]{Ryan W. Pfeifle}
\affiliation{George Mason University, Department of Physics \& Astronomy \\
MS3F3, 4400 University Drive, Fairfax, VA 22030, USA}

\begin{abstract}

We report the discovery of a buried, active supermassive black hole (SMBH) in SDSS J085153.64+392611.76, a bulgeless Seyfert 2 (Sy2) galaxy. Keck near-infrared observations reveal a hidden broad-line region, allowing for the rare case where strong constraints can be placed on both the BH mass and bulge component. Using virial mass estimators, we obtain a BH mass of log($M_{\rm{BH}}/M_{\odot}$) = $6.78 \pm 0.50$. This is one of the only Sy2 active galactic nuclei (AGN) hosted in a bulgeless galaxy with a virial BH mass estimate and could provide important constraints on the formation scenarios of the BH seed population. The lack of a bulge component suggests that the SMBH has grown quiescently, likely caused by secular processes independent of major mergers. In the absence of a detectable bulge component, we find the $M_{\rm{BH}}$--$M_{\rm{stellar}}$ relation to be more reliable than the $M_{\rm{BH}}$--$M_{\rm{bulge}}$ relation. In addition, we detect extended narrow Pa$\alpha$ emission that allows us to create a rotation curve where we see counterrotating gas within the central kiloparsec (kpc). Possible causes of this counterrotation include a galactic bar or disruption of the inner gas by a recent fly-by of a companion galaxy. This in turn could have triggered accretion onto the central SMBH in the current AGN phase.

\end{abstract}


\section{Introduction} \label{sec:intro}

The advent of the discovery that supermassive black holes (SMBHs) lie at the center of virtually all massive galaxies has promoted the idea that these black holes (BHs) play a fundamental role in galaxy formation and evolution \citep[for a review, see][]{Kormendy2013_ARAA}. Well-known relations such as BH mass ($M_{\rm{BH}}$) correlating with stellar velocity dispersion, $M_{\rm{BH}}$--$\sigma$, \citep[eg.,][]{Ferrarese2000,Gebhardt2000,McConnell2013}, and stellar bulge mass, $M_{\rm{BH}}$--$M_{\rm{bulge}}$, \citep[eg.,][]{Marconi2003,Haring2004}, provide a picture that BH growth accompanies central bulge growth. An often-suggested scenario of this interaction involves major mergers that not only fuel BH growth but can also trigger the buildup of the bulge component \citep[eg.,][]{Kauffmann2003,DiMatteo2005,Ellison2011}. Subsequent feedback from the accreting BH (active galactic nucleus [AGN]) can help quench star formation by either expelling gas out of the galaxy \citep[eg.,][]{Kauffmann2000} or by heating the gas in the halo and preventing it from feeding the disk \citep[eg.,][]{Bower2006,Croton2006}. As a result, the galaxy evolves toward the well-defined red sequence. Therefore, galaxy growth has long been thought to be hierarchical, with major mergers providing the necessary conditions for BHs and their host galaxies to reach their observed masses \citep[eg.,][]{Sanders1988,Kauffmann1993}. In addition to major mergers affecting BH growth, stochastic fueling from dense gas clouds reaching the nucleus can also trigger AGN at the low/mid-luminosity regime \citep{Hopkins2014}.

This scenario of BH growth accompanying bulge development through mergers highlights the importance of studying the secular evolution of BHs in galaxies that have had a quiescent merger history.  Unlike bulge-dominated galaxies whose BHs have had accelerated accretion, BHs in likely merger-free galaxies (such as bulgeless galaxies) have grown largely independent of major interactions. Therefore, the mass distribution and occupation fraction of these BHs can provide important clues to the original seed population and secular triggering mechanisms. Additionally, current BH scaling relations lack significant contributions from bulgeless galaxies that can misrepresent the AGN population as a whole. Studying these quiescently grown BHs is thus critical to our understanding of BH growth and their contribution to their host galaxy evolution.

Discoveries of disk-dominated (low bulge-to-total light ratio, B/T) and bulgeless galaxies hosting low- to intermediate-luminosity AGN have been limited, but recent estimates of their BH masses  \citep[eg.,][]{Filippenko2003,Satyapal2007,McAlpine2011,Secrest2012,Reines2013,Simmons2013,Simmons2017} indicate that they can be up to $10^8$ $M_{\odot}$. These findings are starting to show that a central bulge is not a requirement to have a SMBH and that $M_{\rm{BH}}$ in disk-dominated galaxies are likely correlated to the total stellar mass of the galaxy ($M_{\rm{stellar}}$) rather than to the mass of the bulge, $M_{\rm{bulge}}$ \citep{Simmons2017,Martin2018}. In addition, coevolution of BHs and galaxies through merger-free processes, such as disk instabilities and secular growth, has been previously suggested \citep[eg.,][]{Kormendy2004,Greene2010,Schawinksi2011b}, and these processes may be able to grow the central BHs to their typical observed masses \citep{Simmons2013, Martin2018}. This suggests that AGN feedback or perhaps some broader, galaxy-wide process regulates the amount of matter that the BH is allowed to accrete.

However, the number of \textit{purely} bulgeless galaxies with $M_{\rm{BH}}$ estimates in the literature represents a very small fraction of the total bulgeless population, and it is very likely that optical catalogs misidentify or exclude deeply buried AGN in dusty, late-type galaxies. Additional problems arise in verifying the true morphology of the central region, and it is often difficult to rule out the presence of small bulges. This is particularly problematic for Seyfert 1 (Sy1) galaxies, where the bright AGN is in our direct line of sight, compromising the reliability of bulge--disk decompositions employed to measure the total bulge component. While the visible broad-line region (BLR) in these galaxies allows us to obtain estimates of $M_{\rm{BH}}$ through the virial method, the light from the AGN can preclude us from detecting a small bulge component. Sy2 galaxies, on the other hand, allow for much more stringent constraints on the presence of bulges, since their AGN is hidden from our line of sight.  However, for the same reason, $M_{\rm{BH}}$ estimates are more difficult to come by. Several methods have been used in an attempt to detect the `hidden' BLR in Sy2 galaxies, including spectropolarimetry \citep{Antonucci1985} and high S/N near-infrared (NIR) observations where extinction is less severe \citep{Veilleux1997,Lamperti2017}. These studies have revealed that only 10-20\% of Sy2 galaxies show a BLR in the NIR, likely due to strong obscuration.

\begin{figure*}
\centering
\epsscale{1.15}
\plotone{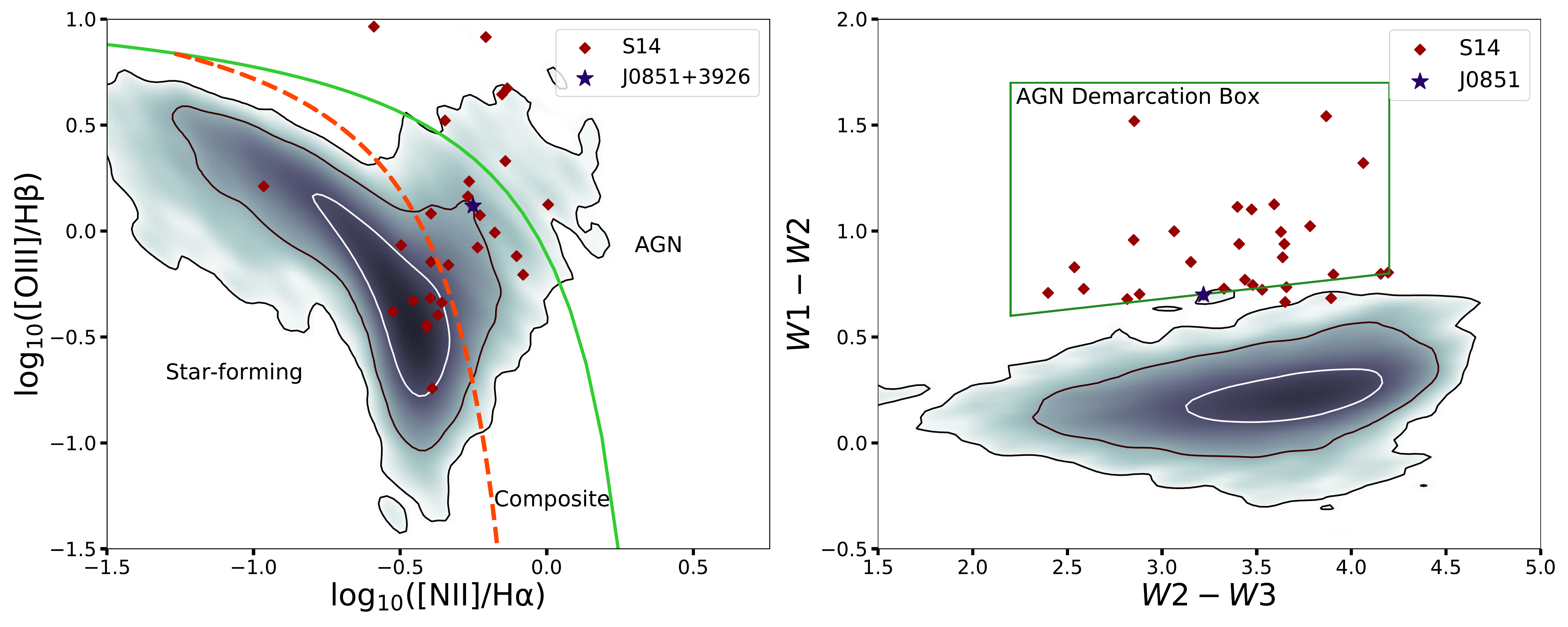}
\caption{BPT (left) and WISE color (right) plots of the bulgeless sample. Red diamonds and the blue star represent the sample selected by S14 and J0851+3926, respectively. The inclusion contours are drawn at $\sigma$ intervals (68\%, 95\%, and 99.5\%). The lines separating the AGN \citep{Kewley2001} and composite \citep{Kauffmann2003} regions from the starforming in the BPT diagram are shown as solid and dashed lines, respectively. The AGN demarcation region is shown as defined in \citet{Jarrett2011}. Note that J0851+3926 falls in the composite region and is on the border of the AGN demarcation box. \label{fig:BPT}}
\end{figure*}

In this article, we present the discovery of a hidden, NIR BLR found in J085153.64+392611.76, hereafter J0851+3926, a spiral galaxy at redshift 0.1296 that shows no signs of a bulge component and is part of a larger study of bulgeless galaxies (T. Bohn et al., in preparation; Fig.~\ref{fig:BPT}). J0851+3926 is listed as `starforming' under the \textit{Subclass} keyword by the Sloan Digital Sky Survey (SDSS), and the SDSS spectrum, although showing composite narrow-line ratios, does not show clear broad Balmer lines. Since there is no AGN contribution at the center, we can put strong constraints on the presence of a bulge using optical photometry. This allows for the rare chance of obtaining a robust BH mass estimate from the NIR broad line while also putting strong constraints on any possible bulge component. In Section \ref{sec:data}, we describe the construction of our sample, observations, and reduction procedure. Section \ref{sec:analysis} presents the results of surface brightness decompositions, the BH mass, intrinsic extinction, and observed gas dynamics. Section \ref{sec:discussion} compares J0851+3926 to other bulgeless galaxies and $M_{\rm{BH}}$--galaxy relations. Additionally, we discuss possible triggering mechanisms of the AGN. We adopt a standard $\Lambda$CDM cosmology with $H_0$ = 70 km s$^{-1}$ Mpc$^{-1}$, $\Omega_M$ = 0.3, and $\Omega_\Lambda$ = 0.7.

\section{Data and Observations} \label{sec:data}
\subsection{Data Selection} \label{subsec:selection}

Since large optical surveys can miss deeply buried AGN, our selection process focuses on infrared (IR) selection techniques. \citet{Sat14}, hereafter S14, selected galaxies believed to host obscured AGN using mid-infrared colors and the AGN selection criteria presented in \citet{Jarrett2011}. S14 suggested that IR indicators could be used to identify optically obscured AGN based on their strong IR colors that separate them from stellar processes. Motivated by these findings, our selection process followed closely to that of S14. To summarize, we formed an initial sample of bulgeless galaxies by drawing from \citet{Simard2011}, who performed bulge-disk decompositions using \textsc{GIM2D} \citep{Simard2002} of 1.12 million galaxies from SDSS DR7. The surface brightness, point-spread function (PSF) convolved bulge--disk decompositions were done in both SDSS \textit{r} and \textit{g} bands. Three different galaxy fitting models were utilized: a bulge ($n_b$ = 4) + disk model, a free-floating S\'ersic index ``bulge'' ($n_b$ = free) + disk model, and a pure S\'ersic model. We used the model with a free-floating bulge index in order to select galaxies with a B/T equal to 0.00. Of the 632,952 galaxies within a redshift of $z<0.2$, only 19,136 have B/T = 0.00 in both \textit{r} and \textit{g} bands. Using fluxes taken from the Portsmouth spectroscopic reanalysis \textit{emissionLinesPort} table \citep{Thomas2013} in SDSS, we constructed a Baldwin, Phillips, $\&$ Terlevich (BPT) diagram \citep{Baldwin1981} using [\ion{O}{3}]$\lambda5007$/H$\beta$ and [\ion{N}{2}]$\lambda6585$/H$\alpha$ line ratios (see Figure \ref{fig:BPT}, left panel). The Portsmouth analysis accounted for stellar absorption features by using the Gas AND Absorption Line Fitting (\textsc{GANDALF}\footnote{\url{https://gandalfcode.github.io}}; \citet{Sarzi2006}) and the penalized Pixel Fitting \citep[\textsc{pPXF}\footnote{\url{https://www-astro.physics.ox.ac.uk/~mxc/software/}};][]{Cappellari2004} routines. By fitting for the stellar absorption features, the fluxes of the hydrogen lines, specifically H$\alpha$ and H$\beta$, increase, causing galaxies to generally shift toward the lower left (i.e., the starforming region) of the BPT diagram. We excluded the 648 galaxies with no registered Portsmouth fluxes, which left only 18,488 bulgeless galaxies in our sample. Following the AGN classification scheme presented in \citet{Kewley2001}, only 143 (0.77\%) galaxies are identified as AGN. However, many spectra can contain contributions from both the AGN and star-forming H II regions. As a result, galaxies hosting relatively weak AGN can fall below this line. \citet{Kauffmann2003} defined a composite region between the AGN and star-forming portions of the diagram, where 950 (5.14\%) galaxies of our sample fall. Galaxies in the composite region are generally believed to have a mixture of AGN and star-forming emission.  However, merger-driven shocks can reproduce AGN-H II emission-line ratios \citep{Rich2014}, so galaxies that fall in this region cannot be definitively classified as AGN without other lines of evidence.

\begin{deluxetable*}{cccccccc}
\caption{Observation Log} 
\label{tab:obs_log}
\tablecolumns{1}
\tablenum{1}
\tablewidth{0pt}
\tablehead{\colhead{Instrument} & \colhead{Date} & \colhead{Seeing} & 
\colhead{Exp. Time} & \colhead{PA} & \colhead{Extraction Aperture} & \colhead{Air Mass} & 
\colhead{Telluric} \\
\colhead{} & \colhead{(YYYY-mm-dd)} & \colhead{(arcsec)} & \colhead{(s)} & \colhead{(deg)} & \colhead{(arcsec)} & \colhead{} & \colhead{}}
\startdata
NIRSPEC & 2018 March 5 & $\sim$ 0.75 & 1920\tablenotemark{a} & 46 & 1.34 & 1.06 & HD 63610 \\
NIRES & 2019 Mar 25 & $\sim$ 0.5 & 1200\tablenotemark{b} & 94 & 1.33 & 1.07 & HD 63610 \\
\enddata
\tablenotetext{a}{8$\times$240 s exposures (two ABBA sets) were taken.}
\tablenotetext{b}{5$\times$240 s exposures were taken.}
\end{deluxetable*}

AGN at low redshift should be considerably redder than inactive galaxies \citep{Stern2012,Assef2013}. Utilizing the Wide-field Infrared Survey Explorer (WISE) All-Sky Data Release \citep{Wright2010}, we obtained WISE band magnitudes \textit{W}1(3.4 $\mu$m), \textit{W}2(4.6 $\mu$m), and \textit{W}3(12 $\mu$m) matched within 1$''$ for our SDSS bulgeless sample. Of the 18,488 SDSS bulgeless galaxies, 18,146 (98.15\%) have registered WISE magnitudes in the required bands. In Figure \ref{fig:BPT} (right panel), we employed the WISE color diagnostic presented in \citet{Jarrett2011}. Here, they define a demarcation zone separating AGNs using \textit{W}1 - \textit{W}2 and \textit{W}2 - \textit{W}3 color cuts. Only 27 members (0.15\%) of our sample fall in the AGN demarcation zone. Using the aforementioned three-band color cut, S14 selected 30 AGN candidates with \textit{W}1 - \textit{W}2 $>$ 0.7 that are most likely to host dominant AGN. These 30 galaxies form our base sample, and one of them, J0851+3926, is the focus of this paper.  The full sample will be investigated in a follow-up paper.

\subsection{NIR Observations and Reductions} \label{subsec:NIR_observations}
NIR spectroscopy of J0851+3926 was obtained on two separate dates: on 2018 March 5 using Keck~II NIRSPEC \citep{McLean1998}, and on 2019 March 25 with Keck II NIRES \citep{Wilson2004}. NIRSPEC is a NIR echelle spectrograph with a wavelength coverage from 0.9 to 5.5 $\mu$m. The NIRSPEC-7 filter was used in low-resolution mode with a cross-dispersion angle of 35.38$^\circ$. This resulted in a wavelength coverage of $\sim$1.8 --- 2.4 $\mu$m. The $42''\times0.76''$ slit was used, and a spectral resolution of $\sim$120 km s$^{-1}$ at the observed wavelength of Pa$\alpha$ was measured with a seeing of $\sim$0.75$''$. Observations throughout the night were done under variable and heavy cloud cover. While telluric and flux standards were observed before and after the science object, the amount of extinction was highly variable, and thus the flux calibration for these data is uncertain. Note that these observations were done before the NIRSPEC upgrade. NIRES is a NIR echelette spectrograph and it has a fixed configuration. The single slit is 18$''\times0.55''$ and the wavelength coverage is set from 0.94 to 2.45 $\mu$m across five orders. There is a small gap in coverage between 1.85 and 1.88 $\mu$m, but this is a region of low atmospheric transmission. The spectral resolution at Pa$\alpha$ was $\sim$85 km s$^{-1}$ and the seeing was typically $\sim$0.5$''$ throughout the night. Observations were done under mostly clear conditions, and so the majority of the analysis was done with the NIRES data. Individual exposures for both sets of observations were 4 minutes each and were done using the standard ABBA nodding. An A0 telluric standard star (with measured magnitudes in $K$, $H$, and $J$ bands) was observed either directly before or after the target galaxy to correct for the atmospheric absorption features.  The total exposure times for NIRSPEC and NIRES were 32 and 20 minutes, respectively. A summary of the NIR observations is shown in Table \ref{tab:obs_log}. 

The data were reduced using two modified pipelines. The first provided flat-fielding and a robust background subtraction by using techniques described in \citet{Kelson2003} and \citet{Becker2009}. In short, this routine maps the 2D science frame and models the sky background before rectification, thus reducing the possibility of artifacts appearing due to the binning of sharp features. The sky subtraction attained with this procedure is excellent, despite the strong OH lines present in the NIR; the procedure is also quite insensitive to cosmic rays and hot pixels and is reliable regardless of skyline intensity.

Rectification, telluric correction, wavelength calibration, and extraction were all done with a slightly modified version of \textsc{REDSPEC}.\footnote{\url{https://www2.keck.hawaii.edu/inst/nirspec/redspec.html}} The sizes of the extracted aperture are listed in Table \ref{tab:obs_log}. The 1D spectrum were then median combined. Flux calibration was done using the telluric star and the Spitzer Science Center unit converter\footnote{\url{http://ssc.spitzer.caltech.edu/warmmission/propkit/pet/magtojy/}} to convert the magnitude of the star to the associated flux in that band. A small corrective factor ($<$ 5$\%$) was introduced owing to the differences between the center of NIR bands and that of the wavelength coverage.

\subsection{X-Ray Observations and Reductions} \label{subsec:Xray_observations}

J0851+3926 was observed for 19.8 ks with the ACIS-S instrument on board the \textit{Chandra} X-ray Observatory on 2020 January 19, with the target centered at the aim point of the ACIS-S3 chip. The data were reduced and analyzed using the Chandra Interactive Analysis of Observations (\textsc{ciao}) software package \citep{fruscione2006} version 4.11 along with version 4.8.2 of the Calibration Database (\textsc{caldb}). A circular aperture of 1.5$''$ in radius was centered on the coordinates of the galaxy nucleus, from which source counts were extracted. The background counts were extracted using a circular aperture of radius 25$''$ and was placed in a nearby area free of other sources. Full (0.3 -- 8~keV), soft (0.3 -- 2~keV), and hard (2 -- 8~keV) counts were extracted from energy-filtered event files using the \textsc{dmextract} package in \textsc{ciao} and error bounds were calculated using Gehrels statistics \citep{gehrels1986}.

\vspace{5mm}

\section{Analysis} \label{sec:analysis}

\subsection{\textsc{GALFIT} Fitting} \label{subsec:Galfit}

J0851+3926 is a Sy2 galaxy with no visible AGN to saturate or blend with a possible bulge. In order to place stringent constraints on the presence of a small bulge, we performed two-dimensional decompositions using \textsc{GALFIT}\footnote{\url{https://users.obs.carnegiescience.edu/peng/work/galfit/galfit.html}} \citep{Peng2002,Peng2010} and ran fits using various combinations of PSF and S\'ersic profiles. The PSF was constructed from the \textit{psField} file provided by SDSS and had a full-width half-maximum (FWHM) of 0.717$''$ (1.65 kpc at the redshift of J0851+3926). Due to the lack of a resolved core component, \textsc{GALFIT} could never converge on a reasonable solution when a PSF or two S\'ersic profiles were used. As expected for a bulgeless galaxy, \textsc{GALFIT} could only properly converge on a solution containing a single S\'ersic profile. As a check, we used various initial index values reflecting a de Vaucouleur bulge ($n=4$), a pseudobulge ($n\sim2$), and an exponential disk ($n=1$). Regardless of the initial values used, \textsc{GALFIT} consistently converged on an index of $n=0.55$.

\begin{figure*}
\gridline{\fig{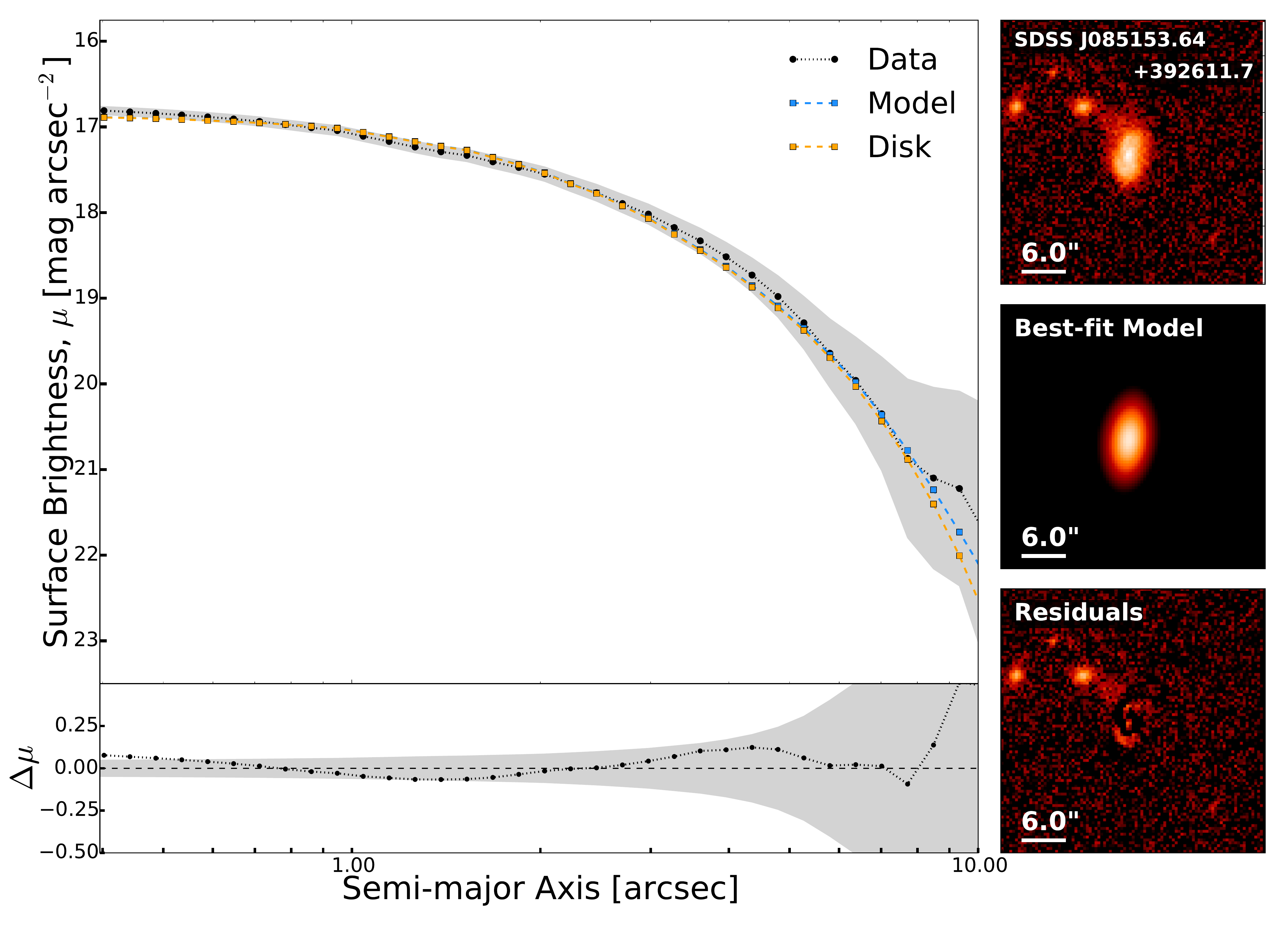}{0.72\textwidth}{(a)}}
\gridline{\fig{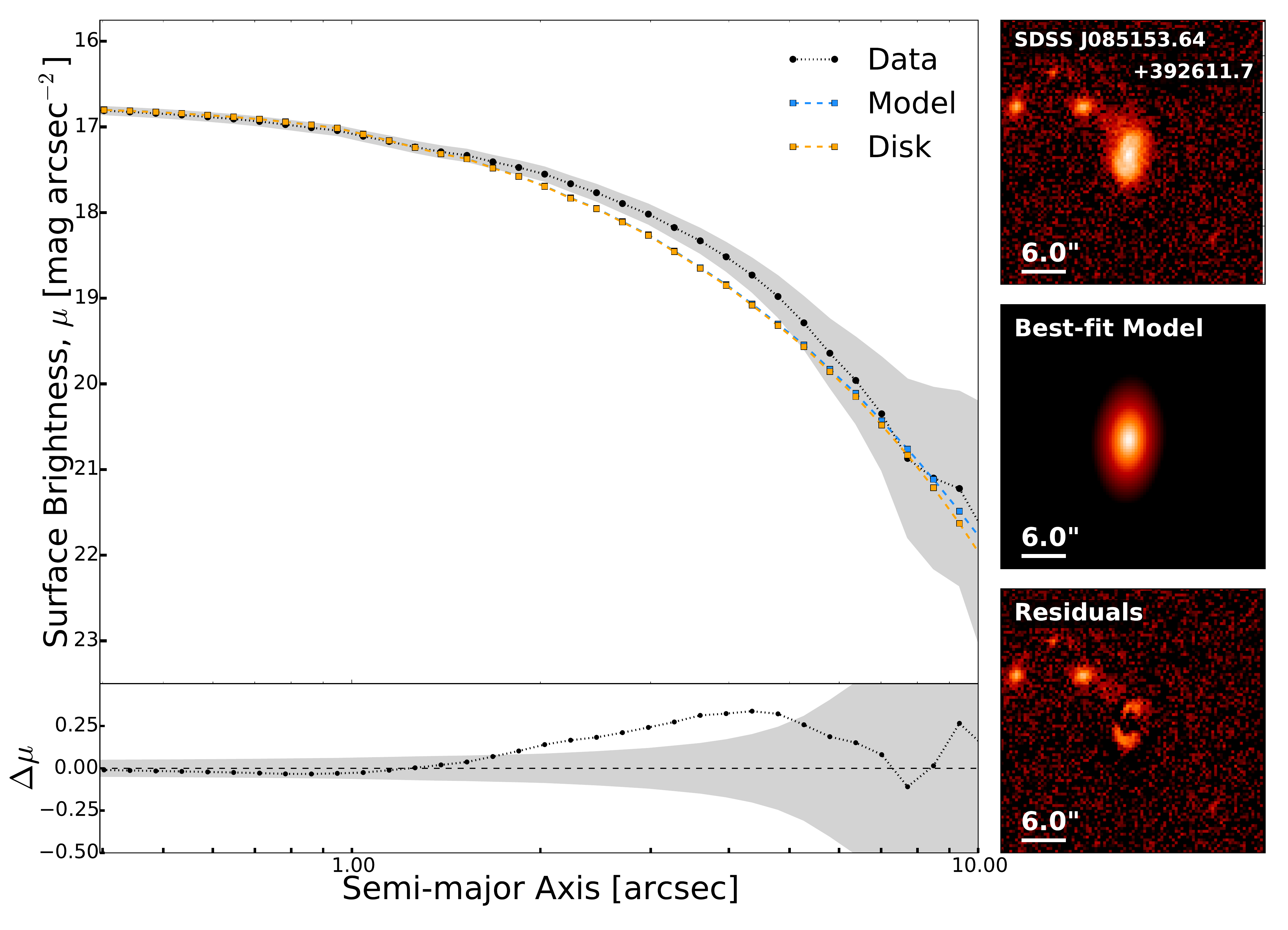}{0.72\textwidth}{(b)}}
\caption{\textsc{GALFIT} decomposition fits to the SDSS image of J0851+3926. Panel (a) is without the outer spirals masked and uses a free S\'ersic index, while panel (b) has the spirals masked and the S\'ersic is fixed at n = 1. In each figure, the fits of the model and disk to the data (black dotted line) are shown in the top left panel as blue and yellow lines, respectively. The model includes both the disk and sky background component. The bottom left panel shows the residuals to the model fit. The shaded gray areas represent the 1-$\sigma$ error. The three right panels show postage stamps of the raw SDSS image, best-fit model, and residuals. \label{fig:galfit}}
\end{figure*}

The fit to the data is presented in Figure \ref{fig:galfit}a, where the best-fit model includes a small background component. The best-fit model matches the data reasonably well out to about 8$''$, where the galaxy ends and the background noise starts to dominate. However, there are some small oscillations in the residuals that are due to spiral arms in the disk of the galaxy, as seen in the 2D residual image. These spiral arms could alter the results of the fitting, particularly in the central region. To factor these out, we tried fitting them using a combination of a power function and Fourier modes, but the resolution and surface brightness of the arms were too low for \textsc{GALFIT} to converge on any solution. We subsequently created a mask using the residuals from the fit and left the central region unmasked. Fitting the galaxy with the mask greatly reduced the residuals at the center and gave a S\'ersic index of $n=0.89$ (see Figure \ref{fig:galfit}b). The central unmasked region within roughly 1.5$''$ is almost perfectly described as a disk with no bulge component.

Although the fits in Figure \ref{fig:galfit} match the data quite well, the possibility of an unresolved bulge or pseudobulge component cannot be dismissed. We obtained an upper limit to the bulge mass by estimating the magnitude of a PSF (i.e., an unresolved component) that can account for the residuals closest to the core in Figure \ref{fig:galfit}a (i.e., the fit without masking the spiral arms). Forcing a PSF that is about 6.5 fainter in magnitude than the total galaxy removed all traces of the central residuals of the original fit. Any PSF brighter than this results in an oversubtraction.  Thus, we take this PSF to be a strong upper limit to the light contribution by an unresolved bulge.  This results in a B/T $\leq$ 0.003, consistent with the value found by \citet{Simard2011}.

To convert this upper limit to a mass, we assumed that the mass-to-light ratio (M/L) is constant throughout the entire galaxy. This is a reasonable assumption since the stellar populations in the disk and bulge components do not differ significantly for disk-dominated galaxies \citep{Graham2001}. We obtained $M_{\rm{stellar}}$ from \citet{Chang2015}, who provide a catalog of stellar masses using SDSS and WISE photometry. Here, SED fitting was performed using both optical and IR imaging to obtain stellar masses. For J0851+3926, a total stellar mass of log($M_{\rm{stellar}}/M_{\odot}$) = 10.61 $\pm$ 0.1 was calculated, which results in an upper limit to the bulge or pseudobulge mass of log($M_{\rm{bulge}}/M_{\odot}$) $\leq$ 8.01.

\subsection{BH Mass of J0851+3926} \label{subsec:J08_mass}

\begin{figure*}
\centering
\epsscale{1.15}
\plotone{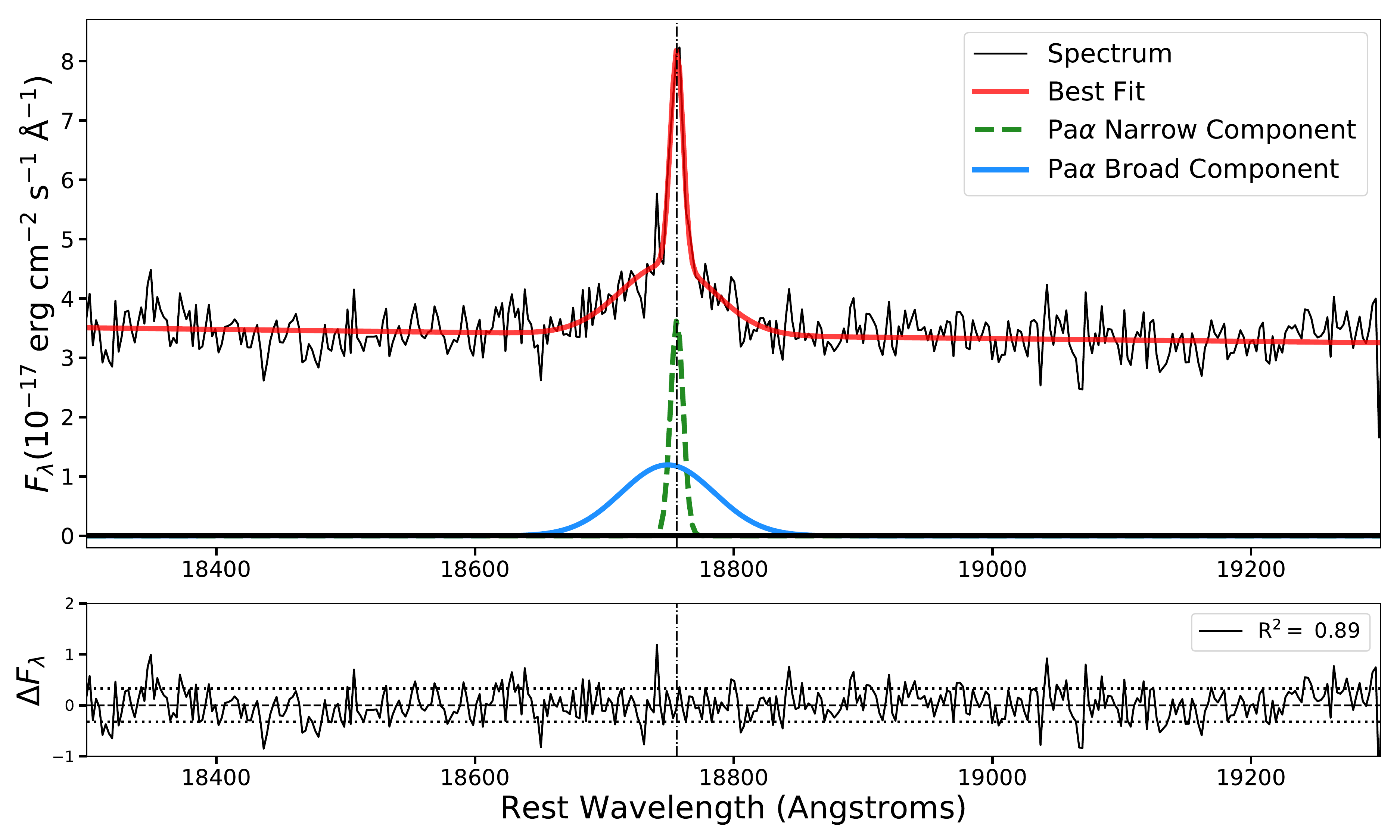}
\caption{The MCMC fit to the NIRES spectrum. In the top panel, Pa$\alpha$ is centered with the best fit plotted over the spectrum. Below the spectrum are the narrow (dotted) and broad (solid) components. The bottom panel plots the residuals, the 1$\sigma$ noise level (horizontal dotted lines), and the $R^2$ value. \label{fig:Pa_broad}}
\end{figure*}

Both the NIRES and NIRSPEC spectra of J0851+3926 clearly show a broad and a narrow component of Pa$\alpha$ (see Figs.~\ref{fig:Pa_broad} and \ref{fig:NIRSPEC_broad}). While broad Pa$\alpha$ is certainly indicative of AGN activity, \citet{Baldassare2016} and other follow-up studies of AGN candidates with broad emission have shown that supernovae (SNe) and other stellar activity can produce similar broad features. Type II SNe \citep{Pritchard2012} and luminous blue variables \citep{Smith2011} are known to produce broad recombination lines up to thousands of kilometers per second. If the broad Pa$\alpha$ observed in J0851+3926 were powered by a SN, the broad emission would have persisted for more than 380 days based on the observation dates of our two sets of spectra (see Table \ref{tab:obs_log}), and this in turn would indicate that the SN would likely be a Type II-P. Using this time scale, we would expect to see other NIR SN features such as \ion{O}{1}, \ion{Mg}{1}, and \ion{Ca}{1} \citep[e.g.,][]{Rho2018}. However, we do not see any of these features in either of our NIR observations. Additionally, we would expect the line profile to change significantly over this time \citep{Rho2018} but our two measurements of the broad Pa$\alpha$ width are consistent with each other, 1489 (NIRSPEC) and 1363 (NIRES) km s$^{-1}$ (see Table \ref{tab:Pa_measurements}).

Another potential origin of a broad line could be powerful outflows powered by star formation.   Broad, symmetric components to emission lines are observed in some starburst galaxies, such as NGC\,1569 \citep{Martin1998,Westmoquette2008,Manzano2019}.  However, these galaxies show broad emission in other lines, particularly in [\ion{O}{3}]$\lambda5007$.  To test whether the broad Pa$\alpha$ in J0851+3926 is powered by an outflow, we fit the optical lines in the SDSS spectrum by using Bayesian AGN Decomposition Analysis for SDSS Spectra (\textsc{BADASS}\footnote{\url{https://github.com/remingtonsexton/BADASS2}}; R.\ Sexton et al. 2020, in preparation), a spectral analysis tool where we included fits to the stellar and \ion{Fe}{2} features, as well as multiple components to emission lines. The code allows the user to test for the presence of outflows by setting various constraints on parameters such as amplitude, width, and velocity offset. All reasonable criteria came back negative for outflows, so we forced a blueshifted, outflow component to the [\ion{O}{3}]$\lambda5007$ fit. We compared the residuals of this forced fit to those fit without outflows and found the residuals to be comparable. This indicates that an outflow component is not needed and that a single gaussian representing narrow-line emission can provide a proper fit to the emission-line profile. Thus, it is likely that outflows that could be affecting the broad line are not present. We conclude that the most likely origin of the observed broad Pa$\alpha$ is the BLR of an AGN. As such, we can use this broad line to estimate BH mass using the virial method.

To obtain a BH mass, a common and reliable method is through the virial relation, defined as 

\begin{equation}
\label{eq:virial}
M_{\rm{BH}} = \mathnormal{f}\frac{V^2 R}{G}
\end{equation}

where $\mathnormal{f}$ is the virial coefficient; $V$ is the velocity of the broad-line gas that is responding to the continuum variations; $R$ is the distance from the broad emission gas to the central continuum source and is equal to $c\tau$ where $\tau$ is the time delay and $c$ is the speed of light; and $G$ is the gravitational constant. The FWHM of the broad emission line, typically seen in H$\alpha$ or H$\beta$ in the optical, can be used for the value of $V$. Here, the width of the broad line stems from the Doppler effect of the gas in the accretion disk revolving around the BH. The value of $R$ is estimated empirically using the optical luminosity of the AGN as a proxy \citep{Kaspi2005,Bentz2013}.

The spectra were fit using \textsc{emcee}, an affine-invariant Markov Chain Monte Carlo (MCMC) ensemble sampler \citep{Foreman-Mackey2013}. A broad component and a narrow component, along with a second-order polynomial for the continuum, were fit simultaneously (see Figure \ref{fig:Pa_broad}). Both the narrow and broad components were treated as Gaussians with amplitude, FWHM, and offset from rest-frame wavelength as free variables. The BH mass was obtained following the estimators presented by \citet{Kim2018}, where they adopted the virial factor log $\mathnormal{f}$ = 0.05 $\pm$ 0.12 derived by \citet{Woo2015}. From \citet{Kim2018}, their Equation 10,

\begin{equation}
\label{eq:2}
\frac{M}{M_{\odot}} = 10^{7.07\pm0.04}\left(\frac{L_{\rm{Pa}\alpha}}{10^{42} \;\rm{erg}\;\rm{s}^{-1}}\right)^{0.49\pm0.06}\left(\frac{\rm{FWHM}_{\rm{Pa}\alpha}}{10^3\;\rm{km}\;\rm{s}^{-1}}\right)^2
\end{equation}

where the FWHM of broad Pa$\alpha$ is the analog to the velocity in the virial mass estimator and $L_{\rm{Pa}\alpha}$ of the broad component is the analog to the distance to the BLR.

Our NIRSPEC observations were done under heavy cloud cover, making it difficult to accurately estimate the degree of extinction. Thus, the flux and BH values are unreliable, but we leave them listed in Table \ref{tab:Pa_measurements} for completeness and show the fit to the spectrum in Appendix \ref{appendix:NIRSPEC_Measure}. For the remainder of the article, we will only use our NIRES data for analysis. From Equation (\ref{eq:2}), our NIRES measurements give a BH mass of ($4.47^{+1.87}_{-1.34}$) $\times \;10^{6}$ $M_{\odot}$ (see Table \ref{tab:Pa_measurements} for relevant values), where the calculated uncertainties come from the random error estimates. Accounting for systematic uncertainties, virial BH mass estimates typically have errors of 0.4 - 0.5 dex \citep[eg.,][]{Shen2013,Reines2015}. We adopt a conservative error estimate of 0.5 dex. For a detailed description on virial mass uncertainties, see \citet{Sexton2019}.

\begin{deluxetable*}{cccccc}[t]
\caption{Pa$\alpha$ Measurements} 
\label{tab:Pa_measurements}
\tablenum{2}
\tablehead{\colhead{Instrument} & \colhead{$\rm{Flux_{Broad}}$} & \colhead{$\rm{Flux_{Broad}}$ (Ext.)\tablenotemark{a}} &  \colhead{$\rm{FWHM_{Broad}}$} & \colhead{${M_{\rm{BH}}}$} &\colhead{${M_{\rm{BH}}}$ (Ext.)\tablenotemark{a}} \\
\colhead{} & \multicolumn{2}{c}{($10^{-15}$ $\rm{erg\; cm^{-2} s^{-1} \AA^{-1}}$)} & \colhead{(km s$^{-1}$)} & \multicolumn{2}{c}{($10^{6}$ $M_{\odot}$)}}
\startdata
NIRSPEC\tablenotemark{b} & 0.306$^{+0.082}_{-0.073}$ & 0.559$^{+0.15}_{-1.33}$ & 1489 $\pm$ 184 & 2.90$^{+2.91}_{-1.56}$ & 3.89$^{+3.64}_{-2.03}$ \\
NIRES & 1.08$^{+0.051}_{-0.050}$ & 1.98$^{+0.100}_{0.098}$ & 1363 $\pm$ 31 & 4.47$^{+1.87}_{-1.34}$ & 6.05$^{+2.23}_{-1.67}$
\enddata
\tablecomments{Listed errors are from random errors in the fitting process.}
\tablenotetext{a}{Extinction-corrected values.}
\tablenotetext{b}{NIRSPEC measurements are ignored owing to the high extinction caused by heavy cloud cover.}
\end{deluxetable*}

\subsection{Extinction} \label{subsec:Extinction}

The presence of broad Pa$\alpha$ and the lack of strong broad lines in the optical imply that there is heavy obscuration present. In order to quantify the extinction toward the BLR, we measured broad hydrogen emission-line ratios and assumed a Cardelli reddening law \citep{Cardelli1989} with an extinction factor $R_V$ = 3.1.  While the observed Pa$\beta$ wavelength fell in a region of strong atmospheric absorption, Pa$\gamma$ emission is observable in the $J$ band, which allowed us to obtain an upper limit to the broad Pa$\gamma$ flux. We also fit the optical SDSS spectrum using \textsc{BADASS}, where fits to the stellar and \ion{Fe}{2} features were included (see R. Sexton et al. 2020, in preparation for further details). Two different models were used to fit the data: one excluding broad components (i.e., only narrow components) and the other including broad components (see Fig. \ref{fig:SDSS_broad} in Appendix \ref{appendix:SDSS_Measure}). No broad H$\beta$ could be properly fit, but the code did converge on a solution to broad H$\alpha$. We compared the fits to H$\alpha$ using the \textit{F}-test: \textit{F} = $(\sigma_{single})^{2}/(\sigma_{double})^{2}$, where $\sigma$ is the standard deviation of the residuals using either single or double gaussian components, for which we obtain \textit{F} = 1.41. Based on this, we cannot say for certain whether adding a broad component is justifiable. A value closer to 2 or 3 would provide convincing evidence that a broad component should be fit. Although the fits do suggest that some broad emission is present, deeper observations will be needed to clear the ambiguity of the broad H$\alpha$ emission.

The observed line ratios from the fits are Pa$\alpha$/Pa$\gamma$ $\geq$ 6.35 and Pa$\alpha$/H$\alpha$ $\geq$ 1.44. Intrinsic line ratios, Pa$\alpha$/Pa$\gamma$ = 3.22 and Pa$\alpha$/H$\alpha$ = 0.10, were obtained from \citet{Dopita2003}, where we assumed an electron density of $n_e$ = 10$^8$ cm$^{-3}$ and temperature $T_e$ = 15,000 K. Using the Cardelli reddening law, we estimated an extinction of $E_{\rm{Pa}\gamma}(B-V)\geq$ 1.13 and $E_{\rm{H}\alpha}(B-V) \geq$ 1.40. Applying $E_{\rm{H}\alpha}(B-V) \geq$ 1.40, the extinction-corrected BH mass is 6.05$^{+2.23}_{-1.67}$ $\times10^{6}$ $M_{\odot}$, which we will use for the remainder of the article (see Table \ref{tab:Pa_measurements} for extinction-corrected values).

This high degree of extinction could explain the lack of other AGN indicators, such as [\ion{Si}{6}]1.963$\mu$m and other coronal lines. To quantify whether extinction could explain their absence, 
we estimate an expected value for the flux of [\ion{Si}{6}] emission based on the observed WISE \textit{W}2(4.6 $\mu$m) flux.  The objects presented by \citet{Muller2018} appear to follow a relation between [\ion{Si}{6}] and \textit{W}2 fluxes as log([\ion{Si}{6}]) = 0.74 $\times$ log($W2$) - 6.6045 (J.\ Cann, private communication). After adjusting for extinction, the expected [\ion{Si}{6}] flux is $\rm{1.05\times 10^{-16}\; erg\; cm^{-2}s^{-1}}$. In order to determine whether if this could be detected, we estimate the flux of a gaussian with an amplitude of 1$\sigma$ of the noise level and a width of the resolution element of the telescope. This resulted in a flux of $\rm{1.12\times 10^{-16}\; erg\; cm^{-2}s^{-1}}$. Thus, any [\ion{Si}{6}] emission will be at most comparable to the noise level. Since [\ion{Si}{6}] is one of the most prominent coronal lines in the NIR, we also do not expect to see other coronal line features.

\subsection{X-Ray Observations} \label{subsec:X_ray_analysis}

The absorbing hydrogen column density ($N_{\rm{H}}$) of Sy2 galaxies is expected to be high, on the order of $\sim$ $\rm{10^{23}\; cm^{-2}}$ \citep{Jaffarian2020}. This is consistent with the unified model, where the active nuclei in Sy2 galaxies are believed to be heavily obscured due to orientation effects. Coupled with the high-extinction estimates calculated in Section \ref{subsec:Extinction}, it is not surprising that we did not detect any statistically significant X-ray emission in the \textit{Chandra} observations. We calculated a $3\sigma$ upper limit on the counts, $\sim6$, and assuming a power-law index of 1.8, the upper limit to the hard X-ray 2 -- 10 keV luminosity, $L_{2-10\ \rm{keV}}$ was estimated to be $1.21\times10^{41}$ erg s$^{-1}$. Using Equation (1) from R. W. Pfeifle et al. (2020, in preparation), a column density can be estimated from $L_{2-10\ \rm{keV}}$ and the WISE $12\rm{\mu m}$ luminosity ($L_{12\rm{\mu m}}$). We found a lower limit line-of-sight column density of $\rm{log}(N_{\rm{H}}/\rm{cm}^{-2}) \geq 24.43$, which suggests that the obscuring region is Compton thick. This estimate of $N_{\rm{H}}$ along with the results presented in \citet{Jaffarian2020} implies an $E(B-V)$ $>$ 1.0, which is consistent with the extinction values calculated in Section \ref{subsec:Extinction}. This heavy obscuration supports the lack of any significant broad emission seen in the optical spectra.

\subsection{Rotation Curve} \label{subsec:Rotation}

\begin{figure*}
\gridline{\fig{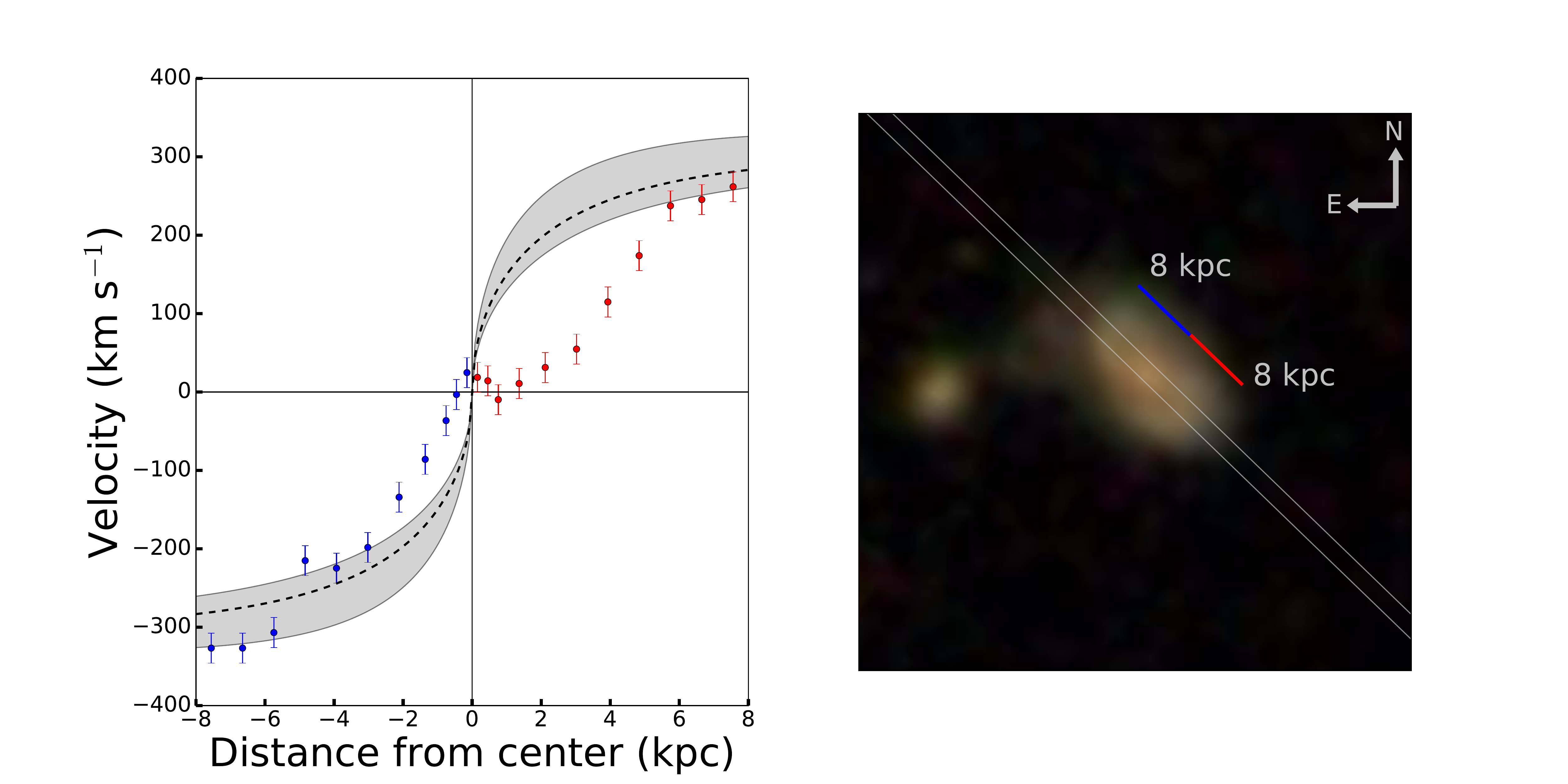}{0.75\textwidth}{(a)}}
\gridline{\fig{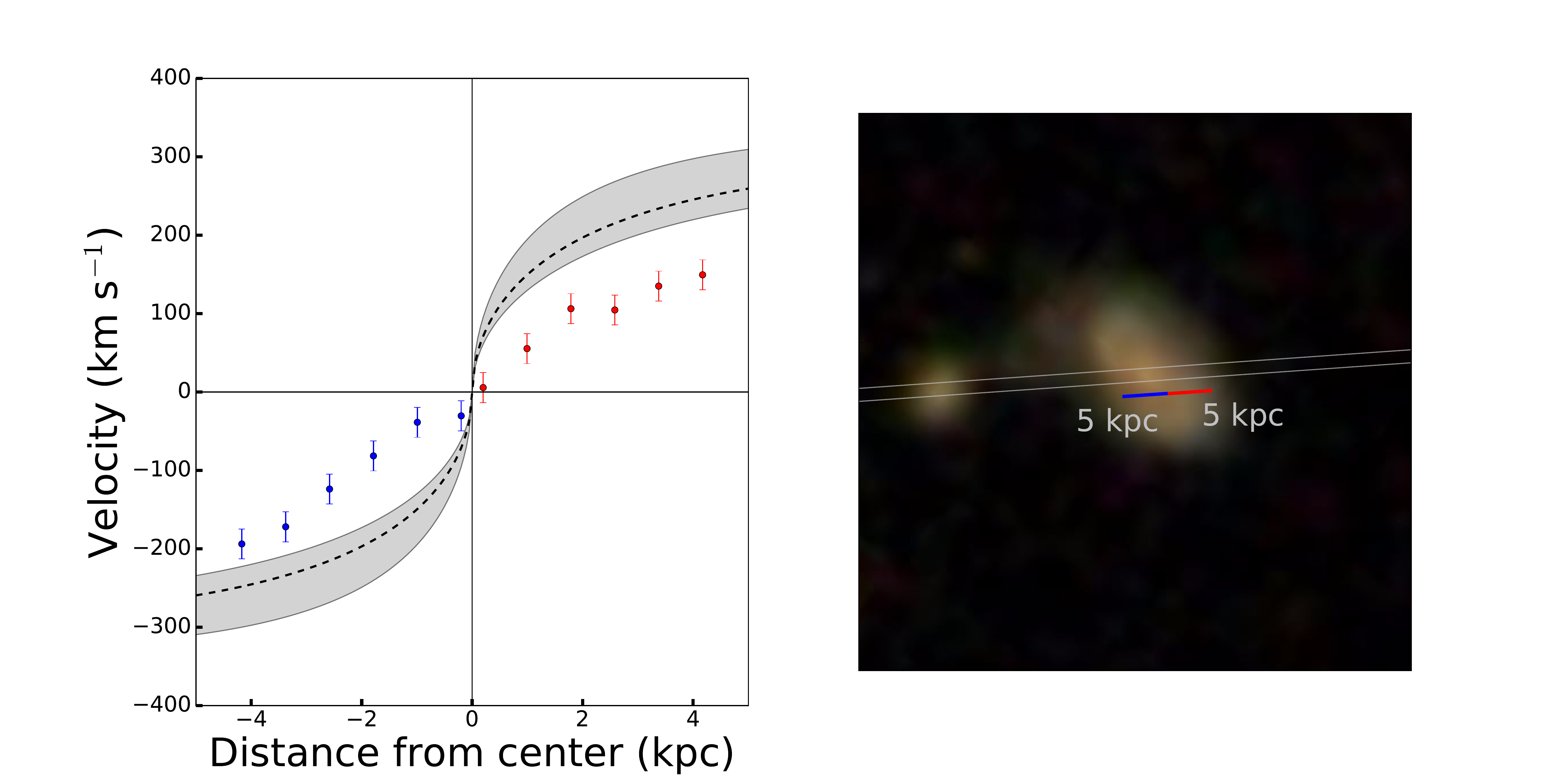}{0.75\textwidth}{(b)}}
\caption{Rotation curves of J0851+3926. A NFW profile is plotted for reference in the left panels. The shaded gray region represents the concentration parameter varying from 8 to 15, with 10 represented as the dotted line. The extended narrow-line velocities of Pa$\alpha$, blue for approaching and red for receding gas, are plotted against their distance from the center in kpc. On the right, the orientation of the slit is shown as a white strip, while the colored lines correspond to the direction of the gas and the distance to which the extended emission is detectable. \label{fig:rot_curve}}
\end{figure*}

Visible in both the NIRSPEC and NIRES 2D spectra is extended narrow Pa$\alpha$ emission that traces the gas in the disk out to about 8 kpc. The spatially resolved narrow emission allows us to construct a rotation curve. Plotted in Figure \ref{fig:rot_curve} are two rotation curves based on different slit orientations, one along the semi-major axis (top) and the other oriented almost perpendicular to that (bottom). For each curve, velocities were measured in either $\sim$0.25 or $\sim$0.5 kpc increments out to the edge of the disk and are color-coded to represent approaching (blue) and receding (red) gas. The fits to the extended emission were done using the same \textsc{emcee} routine as was used in Section \ref{subsec:J08_mass}. The inclination angle is taken from the output of \textsc{GALFIT} (see Figure \ref{fig:galfit}). Plotted in gray is the expected velocity curve using a Navarro-Frenk-White (NFW) dark matter density profile \citep{Navarro1996}. The width of the curve arises by varying the concentration parameter from 8 to 15, with the dotted line representing a value of 10.

As shown in the top panel of Figure \ref{fig:rot_curve}, there appears to be counterrotating gas within the central kpc.  This counterrotation is not apparent when the slit is oriented +60$^\circ$ (bottom panel). The limited extension of the counter-rotation and the orientation of the slit along the central component of the spiral could suggest that a bar is causing the velocity disruption. Another scenario that could explain this counterrotation is a fly-by of a possible companion galaxy. We discuss these scenarios further in Section \ref{subsec:Trigger}.

\section{Discussion} \label{sec:discussion}

The virial method offers one of the most reliable methods of estimating BH masses, and J0851+3928 is one of only a handful of known bulgeless galaxies for which BH mass can be estimated by this method. In the following sections, we first compare the $M_{\rm{BH}}$ of J0851+3928 to those of other bulgeless galaxies, followed by a comparison to a much broader sample that includes all morphological types. We compare $M_{\rm{BH}}$ to both the galactic bulge mass ($M_{\rm{bulge}}$) and total stellar mass of the galaxy ($M_{\rm{stellar}}$). Note that the estimates for $M_{\rm{BH}}$ used here come from methods using the gravitational potential of the SMBH (viral mass estimators) or those using the AGN as the flux source (X-ray estimates). We refrain from using BH masses derived from relations based on galaxy properties, including $M_{\rm{BH}}$-$\sigma$ and $M_{\rm{BH}}$-$\phi$ (spiral arm pitch angle).

\subsection{Comparisons with Other Bulgeless Galaxies}
\label{subsec:bulgeless_comparison}

One of the first examples of an AGN in a bulgeless galaxy in the literature is NGC 4395, a nearby Sy1 galaxy hosting an intermediate-mass BH \citep[IMBH;][]{Filippenko1989,Filippenko2003}. Ultraviolet reverberation mapping has estimated the BH mass to be ($3.6\pm1.1$) $\times \;10^{6}$ M$_{\odot}$ \citep{Peterson2005} and this has been verified by subsequent direct dynamical mass measurements \citep{denBrok2015}. \citet{Jiang2011} report seven broad-line AGN (only 5$\%$ of their sample) where a pure exponential disk provided the best fit, indicating the lack of a bulge component. BH masses in this sample range from $\rm{10^{4.8}}$ to $\rm{10^{6.2}}$ $M_{\odot}$. As noted by the authors, four of these have bar structures and the bright AGN at the center could hide a small bulge. \citet{Simmons2017} provide a large sample (101 galaxies) of type 1 AGN in disk-dominated galaxies with BH masses ranging from $2\times10^{6}$ to $9\times10^{8}$ $M_{\odot}$. However, the mean B/T of this sample is 0.5, possibly caused by the light contribution from the AGN.

Unsurprisingly, only a handful of bulgeless Sy2 galaxies have estimates for $M_{\rm{BH}}$. Most of these observations are confined to IR and X-rays measurements due to the obscuration present in Sy2 galaxies. One such example is NGC 3621, which was first discovered to have AGN activity through IR detections of [\ion{Ne}{5}] at 14 and 24 $\mu$m \citep{Satyapal2007}. Subsequent observations of X-ray emission \citep{Gliozzi2009} and stellar-dynamical modeling of the nuclear star cluster \citep{Barth2009} have placed $M_{\rm{BH}}$ between $4\times10^{3}$ and $3\times10^{6}$ $M_{\odot}$. [\ion{Ne}{5}] detection at 14 $\mu$m was also detected in NGC 4178 \citep{Satyapal2009}. Follow-up X-ray observations \citep{Secrest2012} indicate a  $M_{\rm{BH}}$ between $10^{4}$ and $10^{5}$ M$_{\odot}$. \citet{Shields2008} discovered a low-luminosity IMBH in NGC 1042 with an upper limit to $M_{\rm{BH}}$ calculated at $3\times10^{6}$ $M_{\odot}$ based on the mass of the nuclear star cluster. \citet{McAlpine2011} present two bulgeless galaxies, NGC 3367 (recently identified as a narrow-line Seyfert 1) and NGC 4536. X-ray and [Ne V] detections give estimates of  $M_{\rm{BH}}$ in the range of $10^{5}$ --- $10^{7}$ $M_{\odot}$ and $10^{4}$ --- $10^{6}$ $M_{\odot}$ for NGC 3367 and NGC 4536, respectively. Other bulgeless Sy2 galaxies have only X-ray observations. These include NGC 4561, which has a calculated lower mass limit of $2\times10^{4}$ $M_{\odot}$ \citep{Araya2012}, and NGC 3319, a barred galaxy hosting an IMBH with an estimated upper limit of $3\times10^{5}$ $M_{\odot}$ \citep{Jiang2018}. With $M_{\rm{BH}}$ $\approx$ $10^{6.78}$, J0851+3928 is more massive (in some cases over an order of magnitude) than the other Sy2 bulgeless galaxies listed here.

\begin{figure*}
\centering
\epsscale{0.6}
\plotone{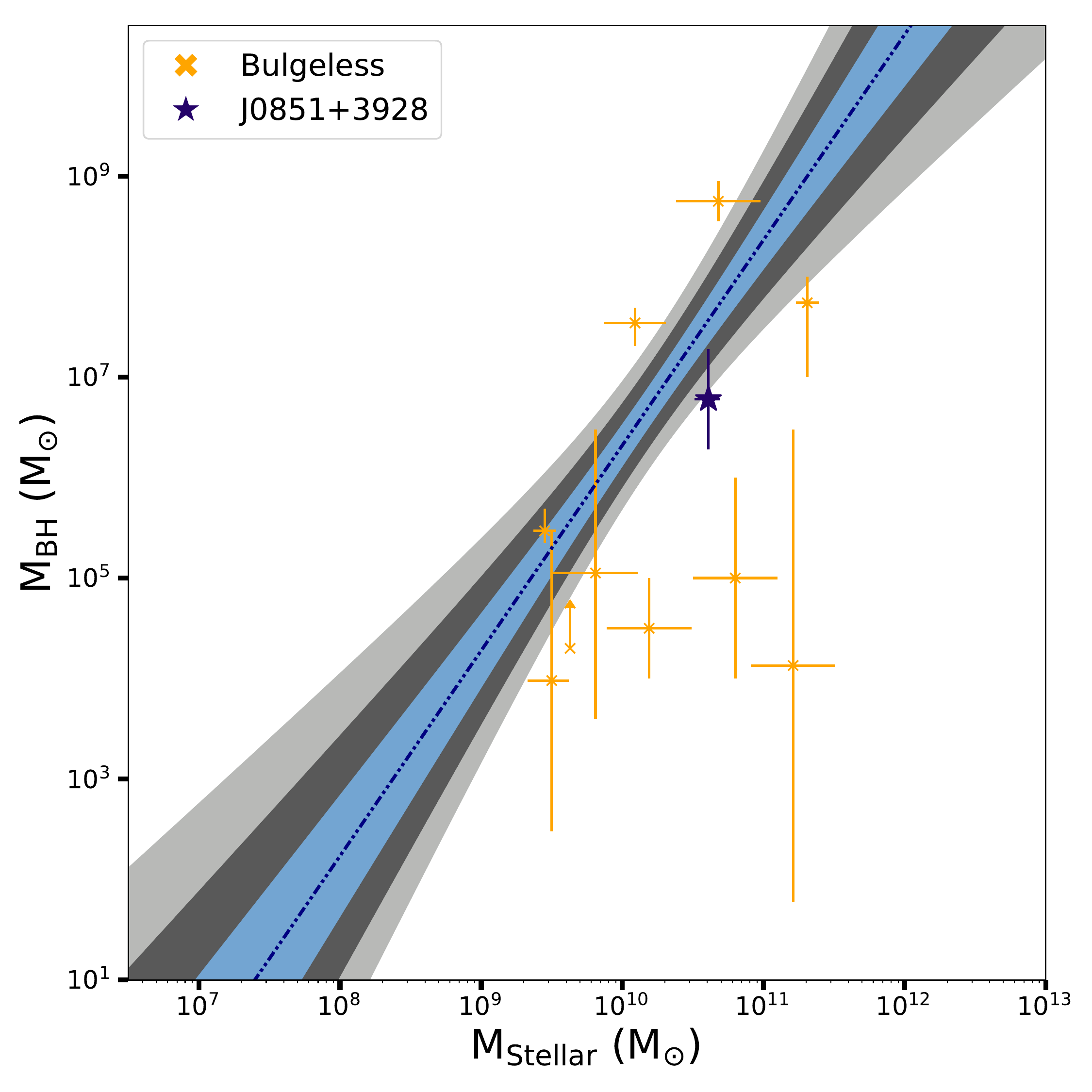}
\caption{$M_{\rm{BH}}$ plotted versus $M_{\rm{stellar}}$ for purely bulgeless galaxies. J0851+3928 is represented as a blue star, and its error bars for stellar mass are comparable to the size of the star symbol. The shaded contours are set at 1$\sigma$ confidence intervals, and the dashed-dotted line represents the line of best fit (excluding J0851+3928).
\label{fig:Bulgeless_stellar_relation}}
\end{figure*}

In Figure \ref{fig:Bulgeless_stellar_relation}, we plot $M_{\rm{BH}}$ vs.\ $M_{\rm{stellar}}$ for the bulgeless sample described above.  Three additional bulgeless galaxies from \citet{Bentz2015online} and \citet{Davis2017} are included (see Tables \ref{tab:BH_relations_reverb} and \ref{tab:BH_relations_dyn} in Appendix \ref{appendix:BH_tables} for mass measurements). In addition, \citet{Rakshit2017} obtained H$\alpha$ line measurements of the NLS1 galaxy NGC 3367, from which we calculated $M_{\rm{BH}}$ using the updated virial mass estimator from \citet{Woo2015} that incorporates the new value of log $\mathnormal{f}$ = 0.05 $\pm$ 0.12 ($\mathnormal{f}$ = 1.12),

\begin{equation}
\label{eq:Ha}
\frac{M}{M_{\odot}} = 10^{5.594\pm0.12}\left(\frac{L_{\rm{H}\alpha}}{10^{42} \;\rm{erg}\;\rm{s}^{-1}}\right)^{0.46}\left(\frac{\rm{FWHM}_{\rm{H}\alpha}}{10^3\;\rm{km}\;\rm{s}^{-1}}\right)^{2.06}
\end{equation}

References for stellar masses and all values are summarized in Appendix \ref{appendix:BH_tables} and Table \ref{tab:BH_relations_bulgeless}. If uncertainties are not given, then they are assumed to be 0.3 dex.

The line of best fit and confidence intervals were calculated using a Bayesian approach with linear regression done by \textsc{emcee} (note that J0851+3928 was not included in this fit). A component of intrinsic scatter was not included due to the significant overlap of the large error bars. The best-fit line is weighed more heavily toward the higher-mass BHs with dynamical mass estimates due to their smaller uncertainties. Fitting only the X-ray observations increased the uncertainty of the slope by almost a factor of 5. Because of the small sample size and large uncertainties, a reliable fit is difficult to make. The virial estimate of $M_{\rm{BH}}$ for J0851+3928 puts it 0.77 dex below the relation but within the scatter of the other $M_{\rm{BH}}$ with virial and dynamical mass estimates. This indicates that the bulgeless BH masses calculated from X-rays are likely lower limits. This is not surprising since many of these galaxies are Sy2, where their high levels of extinction and large column densities can heavily obscure X-ray measurements.

\vspace{5mm}

\subsection{$M_{\rm{BH}}$ Relations} \label{subsec:relations}

\citet{Simmons2013,Simmons2017} have suggested that SMBHs in disk-dominated galaxies are overmassive in the $M_{\rm{BH}}$--$M_{\rm{bulge}}$ relation and seem to outgrow their bulge through secular processes unrelated to major mergers. In addition, they found that these SMBHs follow the $M_{\rm{BH}}$--$M_{\rm{stellar}}$ relation more closely. In order to compare J0851+3928 to these results, we formed an extensive sample that incorporates a range of morphological types with both Sy1 and Sy2 galaxies, including those with pseudobulges. The primary purpose here is to show how J0851+3928 compares to a large sample of galaxies. The following sections describe the sample and papers used, and all measurements are compiled in Appendix \ref{appendix:BH_tables}. The full data set is also available for download.

\subsubsection{BH, Bulge, and Total Stellar Masses} \label{subsubsec:masses_BH_bulge_stellar}

The AGN BH Mass Database \footnote{\url{http://www.astro.gsu.edu/AGNmass/}} \citep{Bentz2015online} provides a compilation of BH masses from reverberation mapping studies. The basic method of reverberation mapping is to monitor variations in the continuum flux and broad emission lines and measure the light-travel time delay between the two. $M_{\rm{BH}}$ are derived from these measurements using the virial relation given by Equation (\ref{eq:virial}).  To properly compare the $M_{\rm{BH}}$ of J0851+3926 calculated using Equation (\ref{eq:2}), we need to adopt a consistent value of $\mathnormal{f}$. Because the reverberation masses in the database are $\sigma$ based, we use log $\mathnormal{f}$ = 0.65 $\pm$ 0.12 ($\mathnormal{f}$ = 4.47) as calibrated in \citet{Woo2015}. $M_{\rm{BH}}$ of 37 galaxies with reliable bulge and stellar masses are listed in Table \ref{tab:BH_relations_reverb} (see Appendix \ref{appendix:BH_tables}). The quoted errors include uncertainties from both the database and from $\mathnormal{f}$, where most of the uncertainty arises.

\citet{Graham2015} compiled data from several different studies and selected the low-mass AGN whose $M_{\rm{BH}}$ are undermassive relative to the $M_{\rm{BH}}$--$M_{\rm{Bulge}}$ relation. These BH masses were calculated using single-epoch virial mass estimators that require the use of the virial coefficient. We use the updated value of $\mathnormal{f}$ ($\mathnormal{f}$ = 1.12) \citep{Woo2015}, as was done in Sections \ref{subsec:J08_mass} and \ref{subsec:bulgeless_comparison}. Due to the need to recalculate all single-epoch mass measurements, we only select AGN from \citet{Graham2015} that have quoted emission-line measurements. The majority of the BH masses were calculated using Equation (\ref{eq:Ha}); however, for Pox 52, where $\lambda L_{\rm{5100}}$ and $\rm{FWHM}_{\rm{H}\beta}$ are reported, we followed the relation derived in \citet{Sexton2019},

\begin{equation}
\begin{split}
\label{eq:Hb}
\frac{M}{M_{\odot}} = 10^{6.867^{+0.155}_{-0.153}}\left(\frac{\lambda\;L_{\rm{5100}}}{10^{44} \;\rm{erg}\;\rm{s}^{-1}}\right)^{0.533^{+0.035}_{-0.033}}\\
\times \left(\frac{\rm{FWHM}_{\rm{H}\beta}}{10^3\;\rm{km}\;\rm{s}^{-1}}\right)^2
\end{split}
\end{equation}

which also incorporates the updated $\mathnormal{f}$ value. Values for $M_{\rm{BH}}$ and the references of the measurements are listed in Table \ref{tab:BH_relations_vir} (see Appendix \ref{appendix:BH_tables}). The errors reported for $M_{\rm{BH}}$ include the quoted uncertainties in $L_{\rm{H}\alpha}$, $\lambda L_{\rm{5100}}$, $\rm{FWHM}_{\rm{H}\alpha}$, $\rm{FWHM}_{\rm{H}\beta}$, and $\mathnormal{f}$.

The rest of the BH masses in our sample were derived using dynamical mass measurements, including stellar dynamics, gas dynamics, stellar orbit motions, and stimulated water maser emission. We compile $M_{\rm{BH}}$ from the following papers: all 44 late-type galaxies in \citet{Davis2017}, 39 early-type galaxies from \citet{Sahu2019}, 37 galaxies from \citet{Savorgnan2016}, and 3 galaxies from \citet{Hu2009}. Masses and uncertainties are quoted from each paper and can be found in Table \ref{tab:BH_relations_dyn} (see Appendix \ref{appendix:BH_tables}).

We also quote $M_{\rm{bulge}}$ and $M_{\rm{stellar}}$ values from the literature. These values are listed in Tables \ref{tab:BH_relations_reverb}--\ref{tab:BH_relations_dyn} and are predominately calculated from color-dependent stellar M/L ratios. The general procedure is to perform 2D bulge/disk decompositions while simultaneously fitting for any structural features such as spiral arms, rings, and bars. Based on the surface brightness profiles, one can obtain apparent and absolute magnitudes from which luminosities can be estimated. With the appropriate M/L ratio, a mass for each component can be calculated. A summary of the methods used to calculate bulge and total stellar mass in each referenced source is presented in Appendix \ref{appendix:BH_tables}. If uncertainties are not specified, then they are assumed to be 0.3 dex.

Lastly, some quoted $M_{\rm{bulge}}$ are greater than their host stellar masses. Although this could naturally come from the use of different methods of fitting or different M/L ratios used, the mass discrepancy could also arise as a result of color differences of the components. For example, in the case of late types in \citet{Bentz2018}, the bulge will tend to be redder and have a higher $V - H$ than the disk. A single $V - H$ that represents the entire galaxy will be bluer, which results in a $M_{\rm{stellar}}$ that is less than $M_{\rm{bulge}}$. The differences, however, are typically within the quoted uncertainties. 

\subsubsection{$M_{\rm{BH}}$-$M_{\rm{bulge}}$ Relations} \label{subsubsec:bulge_relation}

\begin{figure*}
\centering
\epsscale{1.2}
\plotone{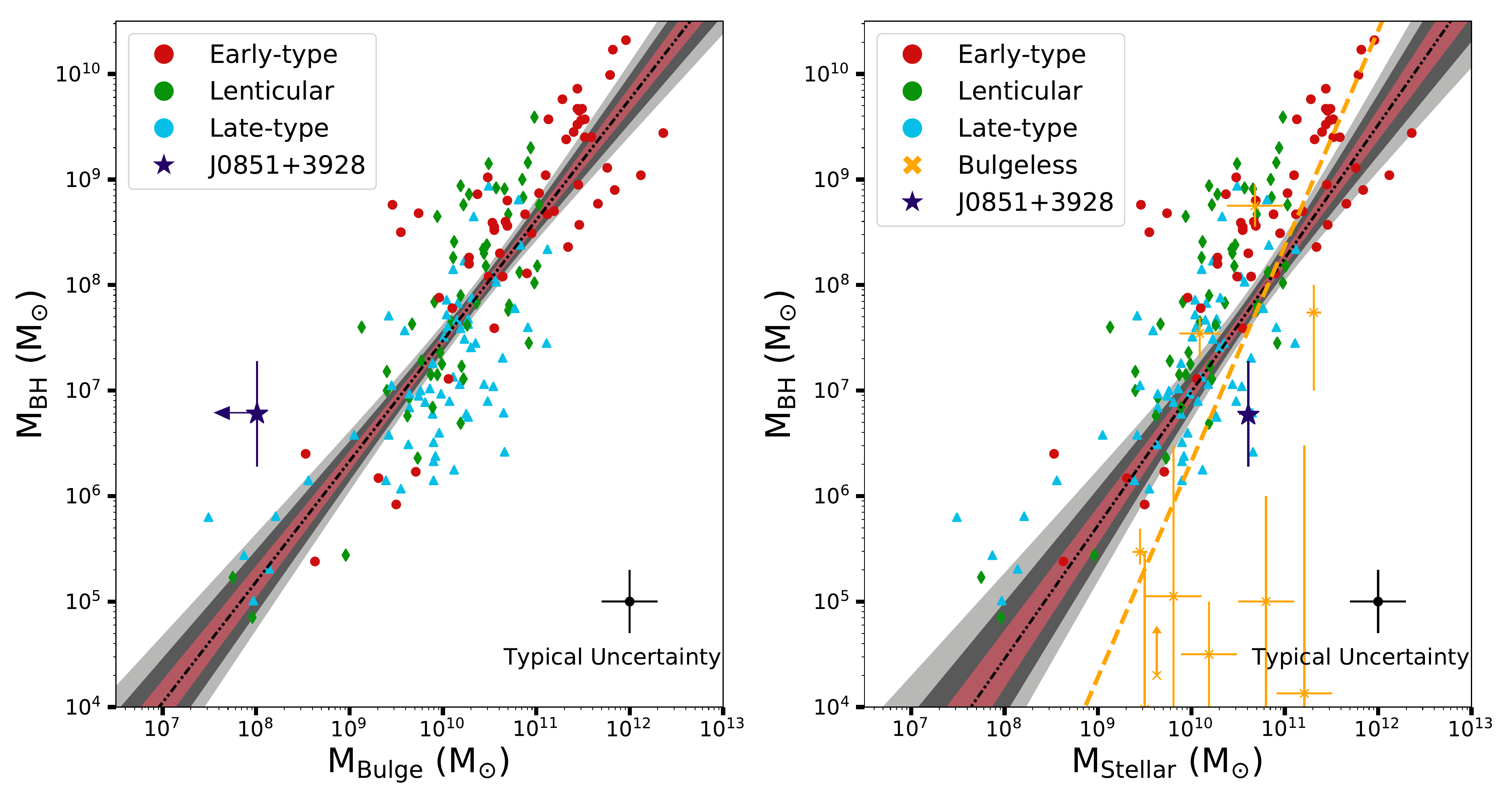}
\caption{$M_{\rm{BH}}$ plotted versus $M_{\rm{bulge}}$ (left) and $M_{\rm{stellar}}$ (right). J0851+3928 is represented as a blue star, and the upper limit to $M_{\rm{bulge}}$ is used here. Error bars for the stellar mass are comparable to the size of the star symbol. The shaded contours are set at 1$\sigma$ confidence intervals, and the black dashed-dotted line represents the line of best fit (excluding J0851+3928 and the bulgeless targets with only X-ray observations). The orange dashed line in the right panel represents the best-fit line to the bulgeless galaxies as is shown in Figure \ref{fig:Bulgeless_stellar_relation}. The full list of individual values and uncertainties are found in Tables \ref{tab:BH_relations_reverb}, \ref{tab:BH_relations_vir} and \ref{tab:BH_relations_dyn} (see Appendix \ref{appendix:BH_tables}). \label{fig:BH_relation}}
\end{figure*}

We first fit $M_{\rm{BH}}$ and $M_{\rm{bulge}}$ data described above using \textsc{emcee} in a similar fashion to what was done in Section \ref{subsec:bulgeless_comparison}; however, a component of intrinsic scatter was included in the fit. This was done since there is not a significant amount of overlap in the error bars. We fit each morphological type individually before fitting the entire sample (note that J0851+3926 was not included). The results of the latter are shown in the left panel of Figure \ref{fig:BH_relation}. Included in the figure are shaded $\sigma$-confidence intervals of the fit, and we find the best-fit relationship to be

\begin{equation}
\label{eq:bulge_relation}
\rm{log}\left(\frac{M_{\rm{BH}}}{M_{\odot}}\right) = (1.14 \pm 0.05) log\left(\frac{M_{\rm{bulge}}}{10^{10} \;{M_{\odot}}}\right) + (7.47 \pm 0.06)
\end{equation}
with an intrinsic scatter of 0.56 dex.

To investigate any potential systematic differences in the various mass calculations used, we also fit each subsample individually. The resulting fits all fall within 1 dex of each other, with only a couple diverging further at the low- and high-mass ends where there is a lack of data to constrain the fit.  The majority of the fits, however, are consistent with the best-fit line for the entire sample, with only slight offsets in the $y$-intercept. We also investigated the morphological dependence of the sample and found that early-type galaxies have a steeper slope (1.13 $\pm$ 0.07) than late-type galaxies (0.81 $\pm$ 0.11). \citet{Davis2019} found that late-type galaxies had a steeper slope; however, our sample includes low-mass early-type galaxies that cause our slope to increase. Interestingly, we investigated the scatter and find that it does not increase significantly ($<$10$\%$) when sampling lower B/T ratios.   This may indicate that even if the mechanism for growing small bulges or pseudobulges is different from that of larger bulges, the resulting bulges/pseudobulges scale with their BHs in a similar manner.

In Section \ref{subsec:Galfit} we obtained an upper limit to the bulge mass of J0851+3928 by forcing the fit of an unresolved component to the central region of the galaxy.  The result indicated that if indeed an unresolved bulge is present in J0851+3928, it is at least 400 times fainter than the disk.  Taken at face value, the upper limit of $M_{\rm{bulge}}$ $\leq$ $10^{8.01}\; \rm{M_{\odot}}$ would indicate that J0851+3928 hosts an overmassive BH compared to what the $M_{\rm{BH}}$ - $M_{\rm{bulge}}$ relation would predict: J0851+3928 is 1.59 dex above the best-fit relation (Fig.~\ref{fig:BH_relation}, left panel), a factor of 2.84 above the scatter. Fitting only late types, we find J0851+3928 to be 1.25 dex above the fit, a factor of 2.40 above the scatter. For galaxies with similar $M_{\rm{BH}}$ (within $10^{6.0}$--$10^{7.0}\; M_{\odot}$), the bulge mass of J0851+3928 is at least 3.50$\sigma$ below the median.

\subsubsection{$M_{\rm{BH}}$-$M_{\rm{stellar}}$ Relation} \label{subsubsec:stellar_relation}

We used the same linear regression method to fit the $M_{\rm{BH}}$ and $M_{\rm{stellar}}$ data as was done for $M_{\rm{BH}}$--$M_{\rm{bulge}}$. The results of the fit to the entire data set are shown in the right panel of Figure \ref{fig:BH_relation} and the best-fit line is

\begin{equation}
\label{eq:stellar_relation}
\rm{log}\left(\frac{M_{\rm{BH}}}{M_{\odot}}\right) = (1.26 \pm 0.09) log\left(\frac{M_{\rm{stellar}}}{10^{10} \;{M_{\odot}}}\right) + (6.99 \pm 0.10)
\end{equation}

with an intrinsic scatter of 0.76, 0.20 dex higher than $M_{\rm{BH}}$ - $M_{\rm{bulge}}$.

Like the $M_{\rm{bulge}}$ relations, the various $M_{\rm{stellar}}$ relations of each subsample do not diverge beyond 1 dex of each other within the low- and high-mass ends. The slope of the early types is steeper than late types, 1.24 $\pm$ 0.10 vs. 1.02 $\pm$ 0.15.  We also find the slope of the $M_{\rm{BH}}$--$M_{\rm{stellar}}$ relation (1.26 $\pm$ 0.10) to be steeper than the $M_{\rm{BH}}$--$M_{\rm{bulge}}$ relation (1.15 $\pm$ 0.06), which is in agreement with \citet{Davis2018} and \citet{Bentz2018}. The fact that $M_{\rm{BH}}$ - $M_{\rm{stellar}}$ is steeper is not surprising since the most massive galaxies hosting the most massive BHs typically have higher B/T flux ratios \citep{Davis2018}. This causes galaxies with higher $M_{\rm{bulge}}$ to be shifted toward the right in the $M_{\rm{BH}}$--$M_{\rm{bulge}}$ relation, thus lowering the steepness of the slope.

Similar to the results found by \citet{Simmons2013,Simmons2017}, J0851+3928 does fall closer to the $M_{\rm{BH}}$ - $M_{\rm{stellar}}$, differing by 0.97 dex, which is a factor of 1.28 below the relation. Also, it is well within the distribution of the late-type galaxies, the most morphologically similar subsample, and only differs by 0.37 dex of the late-type best fit (not shown).

Also plotted in Figure \ref{fig:BH_relation} is the $M_{\rm{BH}}$ - $M_{\rm{stellar}}$ relation for bulgeless galaxies from Figure \ref{fig:Bulgeless_stellar_relation}. The bulgeless X-ray galaxies were not included in the full sample that produced Equation (\ref{eq:stellar_relation}).  We see a decrease in the slope of the full sample, which is likely driven by the handful of galaxies at the low-mass end that have higher $M_{\rm{BH}}$ estimates than the X-ray sources. In addition, the offset of the X-ray sources from the full sample further indicates that these are lower limits to the BH mass. The other bulgeless galaxies with virial estimates for $M_{\rm{BH}}$, including J0851+3928, fall closer to the relation, and J0851+3928 falls well within the scatter of the other late type galaxies. The fact that all four of the bulgeless galaxies with robust estimates fall amongst both early and late-types suggests that perhaps the major BH growth mechanisms in bulgeless galaxies are not all that different. Although the sample size is still too small to make any firm conclusions, it is certainly intriguing how these BHs in bulgeless galaxies grew to supermassive size without going through major merger events. 

\vspace{8mm}

\subsection{Triggering of the AGN} \label{subsec:Trigger}

The existence of a BH on the order of $10^{6.8}$ $M_{\odot}$ in a galaxy with no obvious signatures of a major merger raises the important question of how it has grown to supermassive size. To trigger accretion, an inflow of gas needs to be supplied to the central region. A natural mechanism of this is a galaxy merger in which large quantities of gas can be sent toward the BH. The buildup of the bulge component is thought to accompany major mergers, so a different triggering mechanism likely triggered the AGN activity observed in J0851+3928. In this section, we discuss two possible scenarios of how the current accretion onto the BH may have started: fly-by of a companion galaxy and a galactic bar that can remove angular momentum from the gas.

The SDSS postage stamp (top right panel of Figure \ref{fig:galfit}) shows a small galaxy about 9$''$ away with some low surface brightness emission potentially connecting it to J0851+3928. 
If this galaxy is indeed a companion rather than a close projection, a tidal interaction with J0851+3928 could have disrupted the gas in the disk. If, in addition, this companion were gas-rich, some of the gas could have been accreted by the larger galaxy, possibly explaining the counterrotation observed in Section~\ref{subsec:Rotation}.  In either case, the tidal interaction could have provided the means to remove angular momentum from the gas, thus funneling it onto the central engine.

To investigate whether the small galaxy is a tidal companion, we obtained NIRES spectra to measure its redshift (see Figure \ref{fig:rot_curve}b for slit orientation). Unfortunately, the spectrum did not show any obvious emission or absorption features in any of the NIRES bands.  Thus, the only redshift value available is the photometric redshift (PhotoZ) from SDSS. SDSS reports a PhotoZ = 0.257 $\pm$ 0.0630, compared to PhotoZ = 0.088 $\pm$ 0.0234 and spectroscopic redshift (SpecZ) = 0.129584 ($\pm$ 1.2 $\times$ $10^{-5}$) for J0851+3928. If this redshift is accurate, then the small object is a background galaxy rather than a close companion.  This is consistent with the fact that we find no asymmetries in the residuals of the \textsc{GALFIT} decompositions (see Figure \ref{fig:galfit}), suggesting that the inner disk is largely intact and substantial interaction is unlikely.

\begin{figure}
\begin{center}
\epsscale{1.0}
\plotone{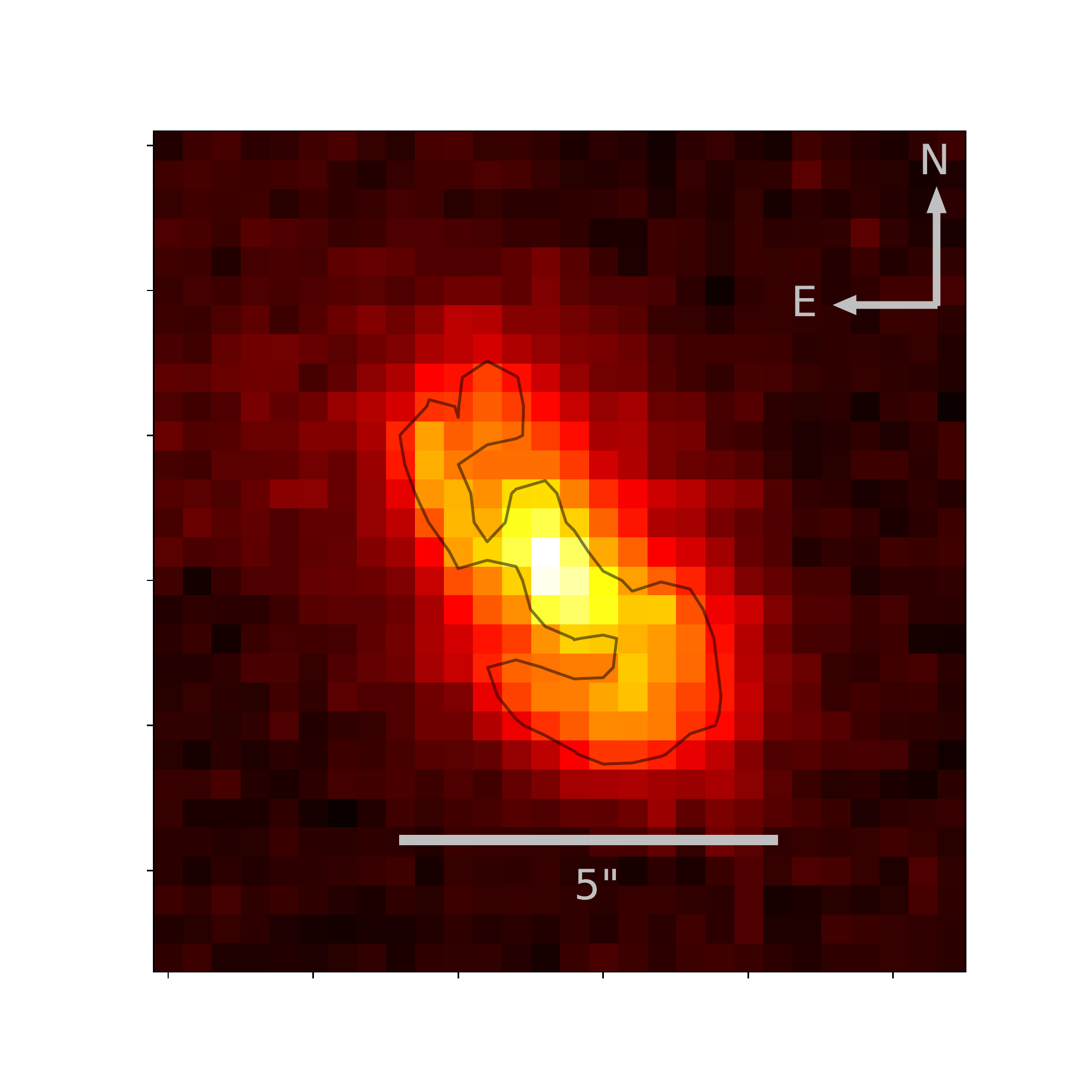}
\caption{Flux plot of J0851+3926. The contour level is arbitrarily  set to bring out the spiral arm structure. \label{fig:contour}}
\end{center}
\end{figure}

Many studies have shown that galactic bars are quite common in nearby spiral galaxies and may play a pivotal role in the secular evolution of AGN \citep[eg.,][]{Eskridge2000,Jogee2005}. Due to the nonaxisymmetric distributions of mass, galactic bars may help drive gas toward the center through gravitational torques that reduce the angular momentum of the gas, thus driving it inward toward parsec scales \citep[eg.,][]{Piner1995,Sheth2002,Wang2012}. Bar formation could have arisen from disk instabilities induced by a close fly-by or cold gas accretion from dark matter filaments \citep{Combes2008}. Through intersecting filaments, \citet{Algorry2014} have shown that a galactic bar can arise due to an inner counterrotating disk or bar. This can occur if accretion along the filaments occurs in different episodes, where the inner bar forms first, followed by the outer disk at a later time. Close inspection of J0851+3928 reveals the possibility of a bar-like structure. Figure \ref{fig:contour} shows a contour, set at an arbitrary level, that emphasizes the spiral arms and possible bar structure. In contrast, no clear bar-like feature is seen in the \textsc{GALFIT} decompositions (see Figure \ref{fig:galfit}). In order to better characterize the central region, higher-resolution data are needed to resolve the central kpc.

\section{Conclusion} \label{sec:Conclusion}

We have obtained NIRSPEC and NIRES NIR spectra and $Chandra$ X-ray observations of SDSS J085153.64+392611.76, a bulgeless, Sy2 galaxy with broad Pa$\alpha$ emission. This offers us the special opportunity to obtain a virial BH mass estimate while also allowing us to put strong constraints on any potential galactic bulge. Using virial mass estimators, we calculated an extinction-corrected BH mass of log($M_{\rm{BH}}/M_{\odot}$) = $6.78 \pm 0.50$. There is some ambiguity to the presence of AGN activity in the SDSS spectrum that showcases that NIR selection techniques could be better suited in selecting and studying AGN that are deeply buried in dust. Our lack of X-ray detection is consistent with this scenario of a heavily obscured AGN and highlights the need for IR spectroscopic observations to uncover hidden BHs in this demographic. Additionally, the lack of a bulge component in J0851+3926 indicates that it is unlikely to have undergone a major merger event and that the central BH has grown to a supermassive size quiescently. Clearly, some secular mechanism, likely independent from mergers, can fuel AGN and grow SMBHs.

We compiled a substantial sample of AGN, including those found in bulgeless galaxies, and find that J0851+3926 falls within the scatter of the $M_{\rm{BH}}$--$M_{\rm{stellar}}$ relation. In addition, the virial mass estimate of the BH mass provides one of the most secure mass estimates of a bulgeless Sy2 galaxy. Since it does not have a bulge component, we find that the $M_{\rm{BH}}$--$M_{\rm{stellar}}$ relation is more reliable for bulgeless galaxies or those with pseudobulges than the $M_{\rm{BH}}$--$M_{\rm{bulge}}$ relation. Obtaining total stellar mass of the galaxy is more straightforward than deconvolving the galaxy into individual components, particularly when the structures are not well resolved.

We also report counterrotation of gas within the central kpc of J0851+3926. Possible causes of this include a potential faint bar that is changing the angular momentum of the gas or a close fly-by of a companion galaxy that disrupted the gas in the disk. Higher-resolution observations will be needed to search for further evidence of a bar and/or traces of tidal interactions.

\acknowledgments

We thank the anonymous referee for their time and helpful comments on this work. We also thank Lisa Prato for her assistance with \textsc{REDSPEC} and George Becker for assisting with the NIR reduction pipeline.

We thank Dr. Greg Doppmann and Dr. Percy Gomez for supporting our Keck observations. We also thank Dr. Laura Sales and Remington Sexton for insightful conversations.

Partial support for this project was provided by the National Science Foundation, under grant No. AST 1817233.

Some of the data presented herein were obtained at the W. M. Keck Observatory, which is operated as a scientific partnership among the California Institute of Technology, the University of California, and the National Aeronautics and Space Administration. The Observatory was made possible by the generous financial support of the W. M. Keck Foundation.

The authors wish to recognize and acknowledge the very significant cultural role and reverence that the summit of Mauna Kea has always had within the indigenous Hawaiian community. We are most fortunate to have the opportunity to conduct observations from this mountain.

Some of the Keck data presented herein were obtained using the UCI Remote Observing Facility, made possible by a generous gift from John and Ruth Ann Evans.

Some of the data and results reported are based on observations made by the Chandra X-ray Observatory.

Funding for the SDSS I/II has been provided by the Alfred P. Sloan Foundation, the Participating Institutions, the National Science Foundation, the U.S. Department of Energy, the National Aeronautics and Space Administration, the Japanese Monbukagakusho, the Max Planck Society, and the Higher Education Funding Council for England. The SDSS website is \url{http://www.sdss.org/}. The  SDSS is managed by the Astrophysical Research Consortium for the Participating  Institutions. The Participating Institutions are the American Museum of Natural History, Astrophysical Institute Potsdam, University of Basel, University of Cambridge, Case Western Reserve University, University of Chicago, Drexel University, Fermilab, the Institute for Advanced Study,  the Japan Participation Group, Johns Hopkins University, the Joint Institute for Nuclear Astrophysics, the Kavli Institute for Particle Astrophysics and Cosmology, the Korean Scientist Group, the Chinese Academy of Sciences (LAMOST), Los Alamos National Laboratory, the Max Planck Institute for Astronomy (MPIA), the Max Planck Institute for Astrophysics (MPA), New Mexico State University, Ohio State University, University of Pittsburgh, University of Portsmouth, Princeton University, the United States Naval Observatory, and the University of Washington.

\vspace{5mm}
\facilities{Chandra, Keck:II (NIRSPEC, NIRES), Sloan.}

\software{\textsc{BADASS} (R. Sexton et al. 2020, in preparation, \url{https://github.com/remingtonsexton/BADASS2}), \textsc{CIAO} (v4.11; software package \citep{fruscione2006}), \textit{emcee} \citep{Foreman-Mackey2013}, \textsc{GALFIT} \citep{Peng2010}, \textsc{GANDALF} \citep{Sarzi2006}, \textsc{pPXF} \citep{Peng2002, Peng2010}, \textsc{PyRAF} (PyRAF is a product of the Space Telescope Science Institute, which is operated by AURA for NASA), \textsc{REDSPEC} (\url{https://www2.keck.hawaii.edu/inst/nirspec/redspec.html})}

\appendix

\section{NIRSPEC Measurements}
\label{appendix:NIRSPEC_Measure}

\begin{figure}[h]
\centering
\epsscale{1.0}
\plotone{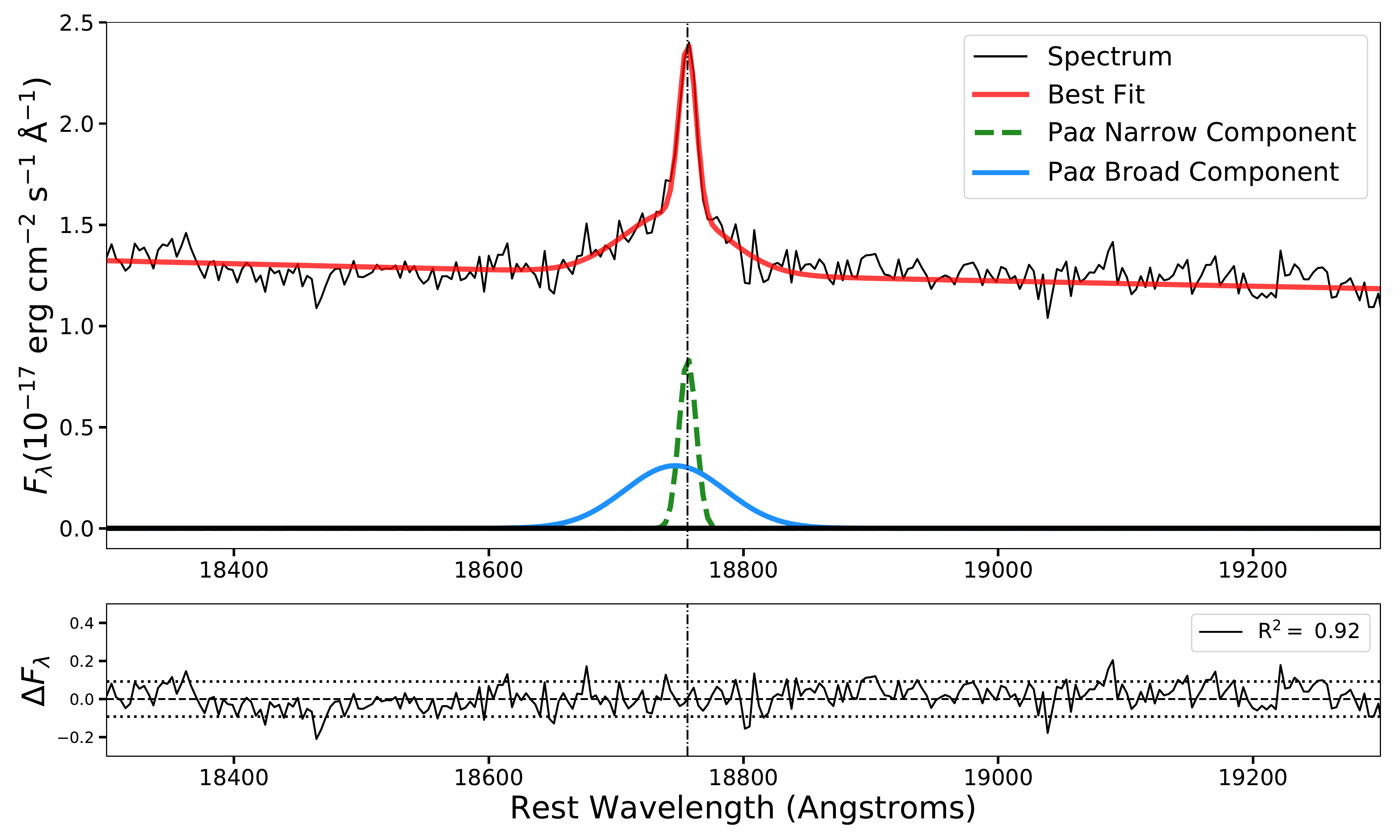}
\caption{The MCMC fit to the NIRSPEC spectrum. In the top panel, Pa$\alpha$ is centered with the best fit plotted over the spectrum. Below the spectrum are the narrow (dotted) and broad (solid) components. The bottom panel plots the residuals, 1$\sigma$ noise level (horizontal dotted lines), and the $R^2$ value. \label{fig:NIRSPEC_broad}}
\end{figure}

Here we show the fit to the NIRSPEC data (see Fig. \ref{fig:NIRSPEC_broad}). As mentioned in Section \ref{subsec:J08_mass}, our NIRSPEC data have a high degree of extinction due to heavy cloud cover, so we do not include the flux measurements in the bulk of our analysis. However, the FWHM of the broad component is still comparable to the NIRES data (see Table \ref{tab:Pa_measurements}) and indicates that the broad emission is due to AGN activity and is not stellar in origin (see Section \ref{subsec:J08_mass} for further details).

\section{SDSS Measurements}
\label{appendix:SDSS_Measure}

\begin{figure}[h]
\centering
\epsscale{1.0}
\plotone{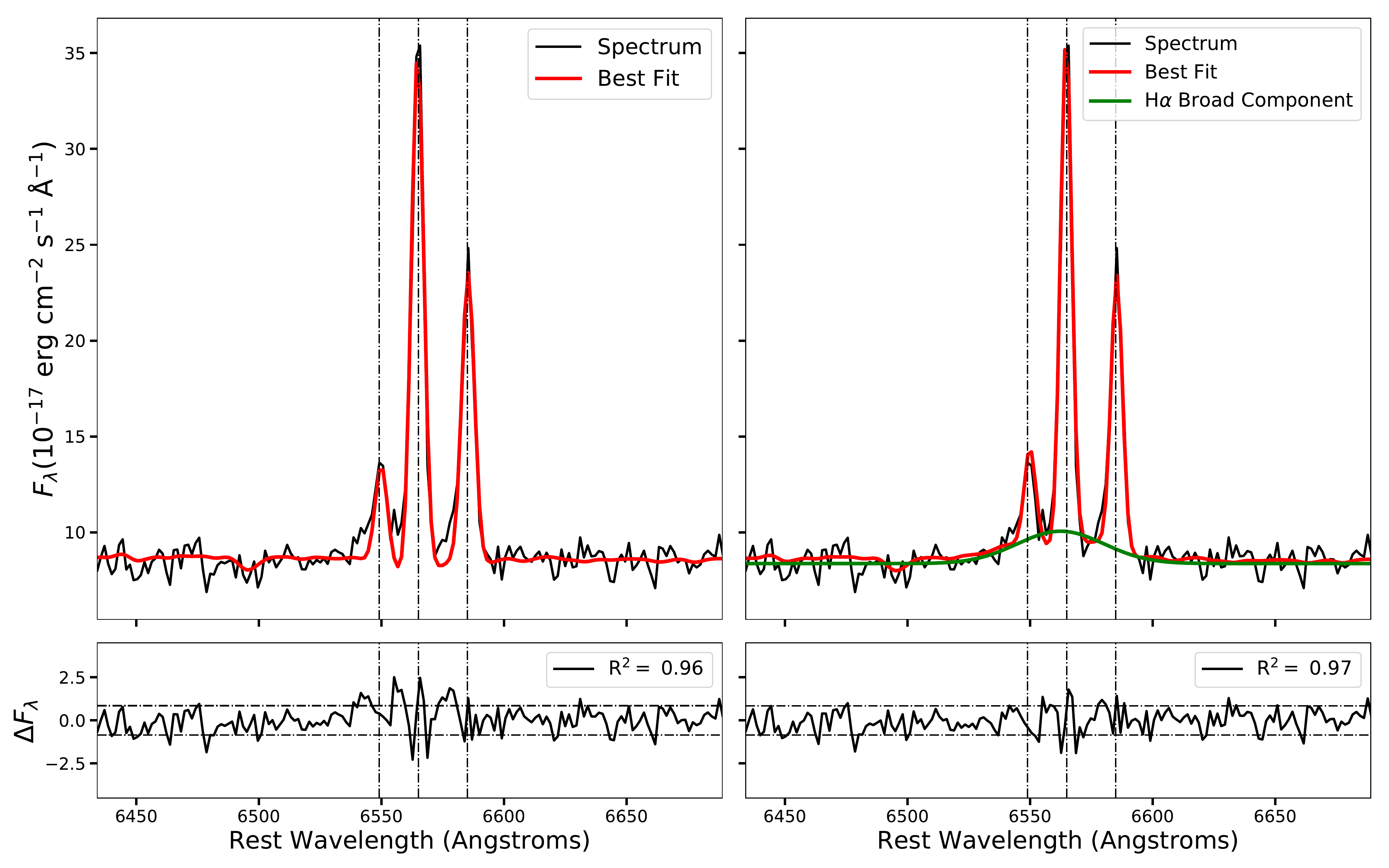}
\caption{The MCMC fit to the SDSS spectrum. In the two top panels, H$\alpha$ and [\ion{N}{2}] $\lambda\lambda$6549, 6585 emission lines are shown, and the best fit to the data is plotted in red. The model on the left uses only single-component gaussian fits to each emission line, while the model on the right adds a broad H$\alpha$ component, plotted in green. The bottom two panels plot the residuals, the 1$\sigma$ noise level (horizontal dotted lines), and the $R^2$ values. \label{fig:SDSS_broad}}
\end{figure}

In this section, we show the \textsc{BADASS} (R. Sexton et al. 2020, in preparation) fits to the SDSS spectrum. We fit two models: one with broad components included (Fig. \ref{fig:SDSS_broad}, right panel) and the other without (Fig. \ref{fig:SDSS_broad}, left panel). The best-fit model (overlaid in red) incorporates both \ion{Fe}{2} emission and stellar absorption from the host galaxy. Outflows were tested for, but none were detected (see R. Sexton et al. 2020, in preparation for further details). Although the full SDSS spectrum was fit, we only show the H$\alpha$ complex since there is substantial absorption in H$\beta$ and no broad H$\beta$ emission is detected. The FWHM of the broad H$\alpha$ emission is $1911.81^{+248.66}_{-212.38}$ km s$^{-1}$, a factor of 1.4 above our NIRES Pa$\alpha$ value. The amplitude is about two times the 1$\sigma$ level of the noise, suggesting that a component could be there.

To compare the fits and quantify whether a broad component is needed, we ran the \textit{F}-test: \textit{F} = $(\sigma_{single})^{2}/(\sigma_{double})^{2}$, where $\sigma$ is the standard deviation of the residuals using either single or double gaussian components. For the region around H$\alpha$, we obtain \textit{F} = 1.41. Although this value suggests that adding a broad component does improve the fit, we cannot say for certain whether a significant broad component exists; a value closer to 2 or 3 is needed to provide more conclusive evidence. It is likely that the broad emission is heavily absorbed and any emission detected is dominated by the noise in the spectrum.

\section{Methods and Measurements for $M_{\rm{BH}}$ Relations}
\label{appendix:BH_tables}

Below is a summary of the methods used to estimate $M_{\rm{bulge}}$ and $M_{\rm{stellar}}$ of our sample described in Sections \ref{subsec:bulgeless_comparison} and \ref{subsec:relations}. The following tables quote the measurements and errors from each reference. Table \ref{tab:BH_relations_bulgeless} lists the bulgeless galaxies discussed in Section \ref{subsec:bulgeless_comparison}. The subsequent tables list mass measurements from Section \ref{subsec:relations} and are categorized based on the method used to estimate $M_{\rm{BH}}$. Note that the $M_{\rm{BH}}$ derived from reverberation mapping and the virial method were recalculated using the updated $\mathnormal{f}$ factor from \citet{Woo2015} (see Section \ref{subsubsec:masses_BH_bulge_stellar}). Lastly, if uncertainties were not listed, then they are assumed to be 0.3 dex. All of the following tables are available for download.

\citet{Fall2018} calculated total stellar masses from $K$-band (2.2 $\mu$m) luminosities using a mass-to-light ratio, $M_{\rm{*}}$/$L_K$, based on $B - V$ colors. To estimate M/L, they used $M_{\rm{*}}$/$L_K$= 0.96($B - V$) + 0.01. For spiral galaxies, which NGC 1024 is classified as, disk stellar masses were calculated separately and summed together with any measured bulge component to get the total stellar mass.

\citet{Davis2018,Davis2019} performed 2D decompositions of 3.6 $\mu$m images from the Spitzer Survey of Stellar Structure in Galaxies, with additional imaging from the \textit{Hubble Space Telescope (HST)} F814W filter and the Two Micron All Sky Survey (2MASS) $K_s$ band (2.2 $\mu$m). M/L ratios of 0.60, 1.88, and 0.62 were used for \textit{Spitzer}, \textit{HST}, and 2MASS data. To account for the contribution of dust emission at 3.6 $\mu$m, a $\sim$ 25$\%$ reduction in the luminosity was included, leading to a $M_{*}/L_{\rm{obs,IRAC1}}$ of 0.453 for dusty galaxies.

\citet{Georgiev2016} obtained total stellar masses using the M/L color relations derived in \citet{Bell2003}. $B$ magnitudes and $B - V$ colors were obtained from \textsc{HyperLeda}\footnote{\url{http://leda.univ-lyon1.fr/}} and were used in log$(M_{*}/L_{B})$ = 1.737$(B-V)$ - 0.942 to obtain $M_{\rm{stellar}}$ for NGC 3319.

\citet{Kelly2012} calculated total stellar masses by first estimating the flux using MAG$\_$AUTO in \textsc{Sextractor}\footnote{\url{http://astroa.physics.metu.edu.tr/MANUALS/sextractor/}}, where the radius of the aperture was set to be 2.5 Kron. M/L ratios were then estimated with SED fitting using \textsc{PEGASE2}\footnote{\url{http://www2.iap.fr/pegase/}} stellar population models.

\citet{McGaugh2014} compiled mass-to-light relations for a number of wavelength bands ($V$, $I$, and 3.6 $\mu$m) from various sources, including those from \citet{Bell2003} and \citet{Into2013}, where $B - V$ colors were used. Magnitudes were taken from \textit{Spitzer} and 2MASS. Values of $M_{\rm{stellar}}$ for NGC 3621 are consistent across all bands (within the assumed error of 0.3 dex), so we use the average of the values given by \citet{Into2013}.

\citet{Hughes2013} utilized the $B - V$ color-dependent relations derived in \citet{Bell2003}. $H$-band (1.65 $\mu$m) luminosities were taken from 2MASS and used in the equation: log$(M_{*}/L_{H})$ = 0.21$(B - V)$ -- 0.059. $B - V$ colors were calculated using either $B$ and $V$ magnitudes from \textsc{GOLDMine}\footnote{\url{http://goldmine.mib.infn.it/}} or morphologically averaged values if observations were not available (taken from \textsc{NED}\footnote{\url{https://ned.ipac.caltech.edu/}}).

\citet{Bentz2018} obtained both optical and NIR imaging of their sample, with the optical data coming from high-resolution medium-band $V$ HST observations and the NIR data from $H$-band images taken at WIYN Observatory. 2D decompositions were done with \textsc{GALFIT} with the higher-resolution HST fits guiding the NIR fit parameters. Masses were calculated using $V-H$ colors and the M/L relation $M_{*}/L_{V}$ = 1.493$(V-H)$ -- 0.681 as derived in \citet{Into2013}. 

\citet{Reines2015} present $M_{\rm{stellar}}$ of a sample of broad-line AGN utilizing SDSS $g$- and $i$-band photometry. A mock AGN spectrum was constructed for each source and then removed to isolate the host luminosity contribution. After correcting for galactic reddening, host galaxy masses were then calculated using log$(M_{*}/L_{i})$ = 1.032$(g - i)$ -- 0.963 \citep{Zibetti2009}. Additional stellar masses for a sample of dwarf galaxies, galaxies with reverberation-mapped AGN, and galaxies with dynamical $M_{\rm{BH}}$ are also provided. The AGN contribution was removed from the dwarf galaxy and reverberation-mapped subsamples, and the stellar mass was obtained in the same way as for the broad-line AGN. For the dynamical BH mass sample and Pox 52, which is not in the SDSS footprint, $B$ and $V$ magnitudes were obtained and stellar masses were calculated using log$(M_{*}/L_{K})$ = 1.176$(B - V)$ -- 1.390 \citep{Zibetti2009}. For Pox 52, a dwarf elliptical, \citet{Barth2004} did not find any indication of a spiral or disk component with \textsc{GALFIT} decompositions, and so we adopt the same value of $M_{\rm{stellar}}$ as $M_{\rm{bulge}}$.

For UM 625, \citet{Graham2015} report a $M_{\rm{bulge}}$ of 5.4 $\times$ $\rm{10^9}$ $M_{\odot}$. This is estimated from a $V$-band bulge magnitude of -19.06 and stellar M/L ratio of 1.6 \citep{Jiang2013}. For $M_{\rm{stellar}}$, we use the value given by \citet{Stern2013}, who obtained masses from SDSS $z$-band photometry. After removal of the AGN contribution, luminosities were converted to masses through a $L\rm{_{[OIII]}}$-dependent M/L ratio. Ratios ranged from 2.6 to 1.7 based on luminosities between $\rm{10^{39}}$ and $\rm{10^{42.5}}$ erg $\rm{s^{-1}}$.

\citet{Chang2015} provide a catalog of stellar masses calculated from spectral energy distribution (SED) fitting of SDSS and WISE photometry. For the two galaxies analyzed, SDSS J004042.10--110957.6 and SDSS J074345.47+480813.5, \citet{Omand2014} have characterized these galaxies as bulge dominated with $r$-band B/T greater than 0.5 (0.53 and 0.58, respectively, when using a de Vaucouleurs model). Using the stellar mass and assuming a constant M/L ratio, we can obtain a rough estimate for $M_{\rm{bulge}}$: log $M\rm{_{bulge}}$ = 9.31 $\pm$ 0.30 for SDSS J004042.10--110957.6 and log $M\rm{_{bulge}}$ = 9.50 $\pm$ 0.30 for SDSS J074345.47+480813.5.

\citet{Schutte2019} expand on the work done by \citet{Reines2015} by calculating $M_{\rm{bulge}}$ of their active dwarf galaxy sample. Optical and IR HST images were run through \textsc{GALFIT} to acquire magnitudes of each component. Magnitudes from HST filters (F606W and F110W) were then converted into SDSS $r$ and $g$ and 2MASS $J$ magnitudes by fitting a wavelength-dependent flux density power law to the HST measurements and then evaluating the fit at the appropriate wavelengths. These new magnitudes were subsequently used in the M/L relation log($M_{*}/L_J$) = 1.398($r - z$) -- 1.271 provided by \citet{Zibetti2009}.

\citet{Savorgnan2016} report bulge luminosities derived from decompositions of 3.6 $\mu$m \textit{Spitzer} images. Individual M/L ratios based on [3.6] -- [4.5] colors were used in the relation log($M_{*}/L_{3.6}$) = 3.98($\pm$0.98)([3.6] -- [4.5]) + 0.13($\pm$0.08) \citep{Meidt2014} to convert luminosities to masses for each galaxy bulge component. 

\citet{Hu2009} analyzed $K$-band images from 2MASS and ran them through \textsc{BUDDA}\footnote{\url{http://www.sc.eso.org/~dgadotti/budda.html}}, a 2D decomposition program, to obtain bulge luminosities. Masses of the bulges were then calculated from either log$(M_{*}/L_{K})$ = 0.135$(B - V)$ -- 0.356 or log$(M_{*}/L_{K})$ = 0.349$(r - i)$ -- 0.336 \citep{Bell2003} where extinction-corrected $B - V$ colors are provided by \textsc{HyperLeda} and the $r - i$ colors are from SDSS. If an AGN component was detected, the central 3$''$ region was removed to avoid contamination from the AGN. When available, we choose the $r - i$ relation since the magnitudes of the bulge effective radius were directly measured using the SDSS images.

\citet{Sahu2019} provide decompositions of early-type galaxies with archived \textit{Spitzer} IRAC 3.6 $\mu$m images, SDSS $r$-band images, or 2MASS $K_s$-band images. \textsc{PROFILER} \citep{Ciambur2015, Ciambur2016} and two \textsc{IRAF} tasks, \textsc{ISOFIT} and \textsc{CMODEL}, were used to model individual galaxy components and obtain magnitudes from which luminosities for the entire galaxy and bulge could be calculated. These luminosities were converted to stellar masses using the following constant stellar M/L ratios for each band: $M_{*}/L_{3.6 \mu m}$ = 0.6, $M_{*}/L_{K_s}$ = 0.7, and $M_{*}/L_{r}$ = 2.8.

For three galaxies, NGC 3414, NGC 4621, and NGC 5846, $M_{\rm{stellar}}$ were obtained from \citet{Dabringhausen2016}. Here, age, color, and luminosity were all treated as parameters in their M/L calculations. Ages of the stellar population for all three galaxies are quoted from \citet{McDermid2015}. $V$-band luminosity and $B - V$ colors came from \textsc{HyperLeda}, and $g - r$ and $g - i$ values came from SDSS. A M/L ratio and $M_{\rm{stellar}}$ were then calculated from these values (see Equation 18 and Table 13 in \citet{Dabringhausen2016}).

\pagebreak

\startlongtable
\begin{deluxetable*}{ccccc}
\tablecaption{Bulgeless Galaxy Sample \label{tab:BH_relations_bulgeless}}
\tablehead{\colhead{Galaxy} & \colhead{log($\frac{M_{\rm{BH}}}{M_{\odot}}$)} & \colhead{Ref} & \colhead{log($\frac{M_{\rm{stellar}}}{M_{\odot}}$)} & \colhead{Ref}} 
\tablenum{3}
\startdata
NGC 1024 & 1.78 - 6.48 & 1 & 11.21 $\pm$ 0.10 & 2 \\
NGC 2748 & 7.54 (+0.15, -0.23) & 3 & 10.09 $\pm$ 0.22 & 4 \\
NGC 3319 & 2.48 - 5.48 & 5 & 9.50 (+0.12, -0.17) & 6 \\
NGC 3367 & 8.75 $\pm$ 0.20 & 7 & 10.68 $\pm$ 0.30 & 8 \\
NGC 3621 & 3.60 - 6.48 & 9 & 9.81 $\pm$ 0.30 & 10 \\
NGC 4178 & 4.0 - 5.0 & 11 & 10.19 $\pm$ 0.30 & 12 \\
NGC 4395 & 5.64 (+0.22, -0.12) & 3 & 9.45 $\pm$ 0.08 & 4 \\
NGC 4536 & 4.0 - 6.0 & 13 & 10.80 $\pm$ 0.30 & 12 \\
NGC 4561 & $>$ 4.30\tablenotemark{a} & 14 & 9.63 $\pm$ 0.30 & 12 \\
NGC 6926 & 7.74 (+0.26, -0.74) & 3 & 11.31 $\pm$ 0.08 & 4
\enddata

\tablenotetext{a}{Lower Limit}
\tablecomments{Columns: (1) Galaxy name. (2) $M_{\rm{BH}}$ estimates from X-ray, IR, or virial measurements, typically lower/upper limits. (3) References for $M_{\rm{BH}}$. (4) Estimates for $M_{\rm{stellar}}$ of the galaxy. (5) References for $M_{\rm{stellar}}$. \\
References: (1) \citep{Shields2008}. (2) \citep{Fall2018}. (3) \citep{Davis2017}. (4) \citep{Davis2018}. (5) \citep{Jiang2018}. (6) \citep{Georgiev2016}. (7) \citep{Rakshit2017}. (8) \citep{Kelly2012}. (9) \citep{Satyapal2007,Barth2009,Gliozzi2009}. (10) \citep{McGaugh2014}. (11) \citep{Secrest2012}. (12) \citep{Hughes2013}. (13) \citep{McAlpine2011}. (14) \citep{Araya2012}.}
\end{deluxetable*}

\startlongtable
\begin{deluxetable*}{ccccccc}
\tablecaption{Galaxy Sample with Reverberation Mapped $M_{\rm{BH}}$ Measurements \label{tab:BH_relations_reverb}}
\tablehead{\colhead{Galaxy} & \colhead{log($\frac{M_{\rm{BH}}}{M_{\odot}}$)} & \colhead{Ref} & \colhead{log($\frac{M_{\rm{bulge}}}{M_{\odot}}$)} & \colhead{Ref} & \colhead{log($\frac{M_{\rm{stellar}}}{M_{\odot}}$)} & \colhead{Ref}}
\tablenum{4}
\startdata
Ark 120 & 8.08 (+0.17, -0.18) & 1 & 10.53 $\pm$ 0.30 & 2 & 10.68 $\pm$ 0.30 & 2 \\
Arp 151 & 6.69 $\pm$ 0.17 & 1 & 10.19 $\pm$ 0.30 & 2 & 10.19 $\pm$ 0.30 & 2 \\
3C 120 & 7.76 $\pm$ 0.16 & 1 & 10.70 $\pm$ 0.30 & 2 & 10.54 $\pm$ 0.30 & 2 \\ 
3C 390.3 & 8.65 (+0.16, -0.17) & 1 & 10.33 $\pm$ 0.30 & 2 & 10.66 $\pm$ 0.30 & 2 \\ 
PG 0026+129 & 8.50 (+0.22, -0.24) & 1 & 9.55 $\pm$ 0.30 & 2 & 9.55 $\pm$ 0.30 & 2 \\
PG 0844+349 & 7.88 (+0.27, -0.35) & 1 & 10.31 $\pm$ 0.30 & 2 & 10.36 $\pm$ 0.30 & 2 \\
PG 1226+023 & 8.86 (+0.20, -0.23) & 1 & 10.37 $\pm$ 0.30 & 2 & 10.37 $\pm$ 0.30 & 2 \\ 
PG 1229+204 & 7.78 (+0.30, -0.34) & 1 & 10.77 $\pm$ 0.30 & 2 & 10.73 $\pm$ 0.30 & 2 \\ 
PG 1307+085 & 8.55 (+0.21, -0.28) & 1 & 10.55 $\pm$ 0.30 & 2 & 10.55 $\pm$ 0.30 & 2 \\ 
PG 1411+442 & 8.56 (+0.25, -0.29) & 1 & 10.69 $\pm$ 0.30 & 2 & 10.69 $\pm$ 0.30 & 2 \\ 
PG 1426+015 & 9.02 (+0.23, -0.28) & 1 & 10.48 $\pm$ 0.30 & 2 & 10.67 $\pm$ 0.30 & 2 \\ 
PG 1613+658 & 8.36 (+0.28, -0.39) & 1 & 11.34 $\pm$ 0.30 & 2 & 11.34 $\pm$ 0.30 & 2 \\ 
PG 1617+175 & 8.68 (+0.20, -0.25) & 1 & 9.74 $\pm$ 0.30 & 2 & 9.74 $\pm$ 0.30 & 2 \\ 
PG 1700+518 & 8.80 (+0.21, -0.22) & 1 & 10.69 $\pm$ 0.30 & 2 & 10.69 $\pm$ 0.30 & 2 \\ 
PG 2130+099 & 7.45 $\pm$ 0.18 & 1 & 11.11 $\pm$ 0.30 & 2 & 10.92 $\pm$ 0.30 & 2 \\ 
SBS 1116+583A & 6.58 (+0.20, -0.21) & 1 & 9.05 $\pm$ 0.30 & 2 & 10.05 $\pm$ 0.30 & 2 \\ 
Zw 229-015 & 6.93 (+0.19, -0.24) & 1 & 9.64 $\pm$ 0.30 & 2 & 9.87 $\pm$ 0.30 & 2 \\ 
Mrk 6 & 8.12 $\pm$ 0.16 & 1 & 10.82 $\pm$ 0.30 & 2 & 10.31 $\pm$ 0.30 & 2 \\ 
Mrk 79 & 7.63 (+0.23, -0.26) & 1 & 10.27 $\pm$ 0.30 & 2 & 10.31 $\pm$ 0.30 & 2 \\ 
Mrk 110 & 7.31 $\pm$ 0.22 & 1 & 10.64 $\pm$ 0.30 & 2 & 10.47 $\pm$ 0.30 & 2 \\ 
Mrk 202 & 6.15 $\pm$ 0.29 & 1 & 9.90 $\pm$ 0.30 & 2 & 9.69 $\pm$ 0.30 & 2 \\ 
Mrk 279 & 7.45 (+0.22, -0.25) & 1 & 10.92 $\pm$ 0.30 & 2 & 10.86 $\pm$ 0.30 & 2 \\ 
Mrk 335 & 7.25 $\pm$ 0.16 & 1 & 9.99 $\pm$ 0.30 & 2 & 9.78 $\pm$ 0.30 & 2 \\ 
Mrk 590 & 7.59 (+0.18, -0.19) & 1 & 10.19 $\pm$ 0.30 & 2 & 11.01 $\pm$ 0.30 & 2 \\ 
Mrk 817 & 7.60 (+0.18, -0.19) & 1 & 10.91 $\pm$ 0.30 & 2 & 10.63 $\pm$ 0.30 & 2 \\ 
Mrk 1310 & 6.23 (+0.19, -0.21) & 1 & 9.71 $\pm$ 0.30 & 2 & 9.53 $\pm$ 0.30 & 2 \\ 
Mrk 1501 & 8.08 (+0.24, -0.29) & 1 & 10.49 $\pm$ 0.30 & 2 & 10.00 $\pm$ 0.30 & 2 \\ 
NGC 3227 & 6.79 (+0.20, -0.23) & 1 & 10.65 $\pm$ 0.30 & 2 & 10.78 $\pm$ 0.30 & 2 \\ 
NGC 3516 & 7.41 (+0.16, -0.18) & 1 & 10.30 $\pm$ 0.32 & 2 & 10.08 $\pm$ 0.32 & 2 \\ 
NGC 4051 & 6.15 (+0.24, -0.28) & 1 & 8.56 $\pm$ 0.32 & 2 & 9.56 $\pm$ 0.32 & 2 \\ 
NGC 4151 & 7.57 $\pm$ 0.17 & 1 & 9.59 $\pm$ 0.32 & 2 & 10.01 $\pm$ 0.32 & 2 \\
NGC 4253 & 6.84 (+0.17, -0.18) & 1 & 9.64 $\pm$ 0.31 & 2 & 9.70 $\pm$ 0.31 & 2 \\ 
NGC 4395 & 5.47 (+0.25, -0.26) & 1 & --- & 3 & 9.45 $\pm$ 0.08 & 4 \\ 
NGC 4593 & 6.90 (+0.20, -0.22) & 1 & 10.48 $\pm$ 0.32 & 2 & 10.40 $\pm$ 0.32 & 2 \\
NGC 4748 & 6.42 (+0.23, -0.30) & 1 & 10.66 $\pm$ 0.30 & 2 & 10.07 $\pm$ 0.30 & 2 \\ 
NGC 6814 & 7.05 $\pm$ 0.18 & 1 & 9.45 $\pm$ 0.35 & 2 & 9.85 $\pm$ 0.35 & 2 \\ 
NGC 7469 & 6.97 $\pm$ 0.17 & 1 & 9.64 $\pm$ 0.30 & 2 & 10.45 $\pm$ 0.30 & 2 \\ 
\enddata
\tablecomments{Columns: (1) Galaxy name. (2) $M_{\rm{BH}}$ estimates from reverberation mapping. (3) References for $M_{\rm{BH}}$. (4) Estimates for $M_{\rm{bulge}}$. (5) References for $M_{\rm{bulge}}$. (6) Estimates for $M_{\rm{stellar}}$ of the galaxy. (7) References for $M_{\rm{stellar}}$.\\
References: (1) \citep{Bentz2015online}. (2) \citep{Bentz2018}. (3) \citep{Davis2019}. (4) \citep{Davis2018}.}
\end{deluxetable*}

\startlongtable
\begin{deluxetable*}{ccccccccc}
\tablecaption{Galaxy Sample with Virial $M_{\rm{BH}}$ Measurements \label{tab:BH_relations_vir}}
\tablenum{5}
\tablehead{\colhead{Galaxy} & \colhead{$\rm{FWHM_{H\alpha}}$\tablenotemark{a}} & \colhead{log($\rm{L_{H\alpha}}$)\tablenotemark{b}} & \colhead{log($\frac{M_{\rm{BH}}}{M_{\odot}}$)} & \colhead{Ref} & \colhead{log($\frac{M_{\rm{bulge}}}{M_{\odot}}$)} & \colhead{Ref} & \colhead{log($\frac{M_{\rm{stellar}}}{M_{\odot}}$)} & \colhead{Ref}}
\startdata
Pox 52 &  765 $\pm$ 30\tablenotemark{c} & 41.64 (+0.10, -0.14)\tablenotemark{d} & 5.38 (+0.16, -0.18) & 1 & 8.63 $\pm$ 0.30 & 5 & 8.63 $\pm$ 0.30 & 5 \\
UM 625 & 1801 $\pm$ 48 & 40.36 $\pm$ 0.01 & 6.36 (+0.15, -0.14) & 2 & 9.73 $\pm$ 0.30 & 6 & 10.00 $\pm$ 0.30 & 9 \\
SDSS J004042.10-110957.6 & 2240 $\pm$ 224 & 39.53 $\pm$ 0.05 & 6.17 (+0.24, -0.23) & 3 & 9.31 $\pm$ 0.30 & 7 & 9.59 $\pm$ 0.10 & 10 \\
SDSS J074345.47+480813.5 & 1450 $\pm$ 145 & 39.81 $\pm$ 0.05 & 5.92 (+0.23, -0.24) & 3 & 9.50 $\pm$ 0.30 & 7 & 9.74 $\pm$ 0.09 & 10 \\
SDSS J024656.39-003304.8 & 1577 $\pm$ 158 & 39.38 (+0.06, -0.08) & 5.81 (+0.22, -0.27) & 4 & 8.21 $\pm$ 0.30 & 8 & 9.45 $\pm$ 0.30 & 5 \\
SDSS J090613.75+561015.5 & 703 $\pm$ 70 & 40.15 $\pm$ 0.02 & 5.44 (+0.20, -0.24) & 4 & 8.96 $\pm$ 0.30 & 8 & 9.30 $\pm$ 0.30 & 5 \\
SDSS J095418.15+471725.1 & 636 $\pm$ 64 & 39.41 $\pm$ 0.06 & 5.01 (+0.22, -0.26) & 4 & 7.97 $\pm$ 0.30 & 8 & 9.24 $\pm$ 0.30 & 5 \\
SDSS J144012.70+024743.5 & 747 $\pm$ 75 & 39.73 (+0.05, -0.06) & 5.31 (+0.21, -0.26) & 4 & 8.14 $\pm$ 0.30 & 8 & 9.30 $\pm$ 0.30 & 5 \\
SDSS J085125.81+393541.7 & 894 $\pm$ 89 & 39.67 (+0.05, -0.06) & 5.44 (+0.21, -0.26) & 4 & 7.87 $\pm$ 0.30 & 8 & 9.12 $\pm$ 0.30 & 5 \\
SDSS J152637.36+065941.6 & 1043 $\pm$ 104 & 40.16 $\pm$ 0.02 & 5.80 (+0.20, -0.24) & 4 & 7.49 $\pm$ 0.30 & 8 & 9.36 $\pm$ 0.30 & 5 \\
SDSS J160531.84+174826.1 & 792 $\pm$ 79 & 39.45 $\pm$ 0.05 & 5.23 (+0.21, -0.26) & 4 & 7.75 $\pm$ 0.30 & 8 & 9.36 $\pm$ 0.30 & 5 \\  
\enddata
\tablenotetext{a}{FWHM are in units of km s$^{-1}$.}
\tablenotetext{b}{Luminosities are in units of erg s$^{-1}$.}
\tablenotetext{c}{FWHM of $\rm{H\beta}$.}
\tablenotetext{d}{log($\rm{\lambda L_{5100}}$).}
\tablecomments{Columns: (1) Galaxy Name. (2) FWM of the broad $\rm{H\alpha}$ (or $\rm{H\beta}$ for Pox 52) emission line. (3) Luminosity of broad $\rm{H\alpha}$ (or $\rm{\lambda L_{5100}}$ for Pox 52). (3) $M_{\rm{BH}}$ estimates from virial measurements. (3) References for $M_{\rm{BH}}$. (4) Estimates for $M_{\rm{bulge}}$. (5) References for $M_{\rm{bulge}}$. (6) Estimates for $M_{\rm{stellar}}$ of the galaxy. (7) References for $M_{\rm{stellar}}$.\\
References: (1) \citep{Thornton2008}. (2) \citep{Jiang2013}. (3) \citep{Yuan2014}. (4) \citep{Reines2013}. (5) \citep{Reines2015}. (6) \citep{Graham2015}. (7) \citep{Omand2014} (8) \citep{Schutte2019}. (9) \citep{Stern2013}. (10) \citep{Chang2015}.}
\end{deluxetable*}

\startlongtable
\begin{deluxetable*}{ccccccc}
\tablecaption{Galaxy Sample with Dynamical $M_{\rm{BH}}$ Measurements \label{tab:BH_relations_dyn}}
\tablehead{\colhead{Galaxy} & \colhead{log($\frac{M_{\rm{BH}}}{M_{\odot}}$)} & \colhead{Ref} & \colhead{log($\frac{M_{\rm{bulge}}}{M_{\odot}}$)} & \colhead{Ref} & \colhead{log($\frac{M_{\rm{stellar}}}{M_{\odot}}$)} & \colhead{Ref}} 
\tablenum{6}
\startdata
Milky Way & 6.60 $\pm$ 0.02 & 1 & 9.96 $\pm$ 0.05 & 5 & 10.78 $\pm$ 0.10 & 6 \\
Circinus  & 6.25 (+0.07, -0.08) & 1 & 10.12 $\pm$ 0.20 & 5 & 10.62 $\pm$ 0.18 & 6 \\
Cygnus A & 9.44 (+0.11, -0.14) & 1 & 12.36 $\pm$ 0.20 & 5 & 12.38 $\pm$ 0.20 & 6 \\
ESO 558-G009 & 7.26 (+0.03, -0.04) & 1 & 9.89 $\pm$ 0.11 & 5 & 11.03 $\pm$ 0.10 & 6 \\
IC 1459 & 9.38 (+0.15, -0.23) & 2 & 11.32 $\pm$ 0.15 & 3 & 11.28 $\pm$ 0.30 & 7 \\
IC 4296 & 9.04 (+0.07, -0.09) & 2 & 12.12 $\pm$ 0.15 & 3 & 11.45 $\pm$ 0.30 & 7 \\
IC 2560 & 6.49 (+0.08, -0.10) & 1 & 9.63 $\pm$ 0.39 & 5 & 10.66 $\pm$ 0.37 & 6 \\
PGC 49940 & 9.59 (+0.05, -0.06) & 3 & 10.98 $\pm$ 0.15 & 3 & 11.54 $\pm$ 0.12 & 7 \\ 
SDSS J043703.67+245606.8 & 6.51 (+0.04, -0.05) & 1 & 9.90 $\pm$ 0.20 & 5 & 10.97 $\pm$ 0.10 & 6 \\ 
Mrk 1029 & 6.33 (+0.10, -0.13) & 1 & 9.90 $\pm$ 0.11 & 5 & 10.66 $\pm$ 0.09 & 6 \\ 
NGC 221 & 6.40 (+0.08, -0.10) & 3 & 8.53 $\pm$ 0.15 & 3 & 8.77 $\pm$ 0.30 & 7 \\ 
NGC 224 & 8.15 (+0.22, -0.10) & 1 & 10.11 $\pm$ 0.09 & 5 & 10.88 $\pm$ 0.10 & 6 \\ 
NGC 253 & 7.00 $\pm$ 0.30 & 1 & 9.76 $\pm$ 0.09 & 5 & 10.71 $\pm$ 0.08 & 6 \\ 
NGC 307 & 8.34 $\pm$ 0.13 & 4 & 10.43 $\pm$ 0.33 & 4 & 10.76 $\pm$ 0.12 & 4 \\ 
NGC 404 & 4.85 $\pm$ 0.13 & 4 & 7.96 $\pm$ 0.27 & 4 & 9.12 $\pm$ 0.12 & 4 \\ 
NGC 524 & 8.92 $\pm$ 0.10 & 4 & 10.57 $\pm$ 0.26 & 4 & 11.07 $\pm$ 0.12 & 4 \\ 
NGC 821 & 7.59 (+0.22, -0.11) & 2 & 10.55 $\pm$ 0.15 & 3 & 10.66 $\pm$ 0.30 & 7 \\ 
NGC 1023 & 7.62 $\pm$ 0.04 & 2 & 10.26 $\pm$ 0.15 & 3 & 10.63 $\pm$ 0.30 & 7 \\ 
NGC 1068 & 6.75 $\pm$ 0.02 & 1 & 10.27 $\pm$ 0.24 & 5 & 10.78 $\pm$ 0.18 & 6 \\ 
NGC 1097 & 8.38 $\pm$ 0.03 & 1 & 10.83 $\pm$ 0.20 & 5 & 11.40 $\pm$ 0.10 & 6 \\ 
NGC 1194 & 7.81 $\pm$ 0.04 & 4 & 10.71 $\pm$ 0.33 & 4 & 10.94 $\pm$ 0.12 & 4 \\ 
NGC 1275 & 8.90 $\pm$ 0.20 & 4 & 11.84 $\pm$ 0.26 & 4 & 11.88 $\pm$ 0.12 & 4 \\ 
NGC 1300 & 7.71 (+0.17, -0.12) & 1 & 9.42 $\pm$ 0.25 & 5 & 10.30 $\pm$ 0.17 & 6 \\ 
NGC 1316 & 8.18 (+0.18, -0.33) & 2 & 11.01 $\pm$ 0.15 & 3 & 11.48 $\pm$ 0.30 & 7\\ 
NGC 1320 & 6.78 (+0.16, -0.26) & 1 & 10.25 $\pm$ 0.40 & 5 & 10.58 $\pm$ 0.40 & 6 \\ 
NGC 1332 & 9.16 (+0.06, -0.07) & 2 & 10.91 (+0.26, -0.35) & 2 & 10.92 $\pm$ 0.30 & 7 \\ 
NGC 1374 & 8.76 $\pm$ 0.05 & 4 & 10.22 $\pm$ 0.26 & 4 & 10.52 $\pm$ 0.12 & 4 \\ 
NGC 1398 & 8.03 $\pm$ 0.08 & 1 & 10.57 $\pm$ 0.20 & 5 & 11.25 $\pm$ 0.18 & 6 \\ 
NGC 1399 & 8.67 (+0.05, -0.06) & 2 & 11.12 $\pm$ 0.15 & 3 & 11.17 $\pm$ 0.30 & 7 \\ 
NGC 1407 & 9.65 $\pm$ 0.08 & 4 & 11.46 $\pm$ 0.27 & 4 & 11.52 $\pm$ 0.12 & 4 \\ 
NGC 1550 & 9.57 $\pm$ 0.06 & 4 & 11.13 $\pm$ 0.12 & 4 & 11.13 $\pm$ 0.12 & 4 \\ 
NGC 1600 & 10.23 $\pm$ 0.05 & 4 & 11.82 $\pm$ 0.12 & 4 & 11.82 $\pm$ 0.12 & 4 \\ 
NGC 2273 & 6.97 $\pm$ 0.03 & 1 & 9.98 $\pm$ 0.20 & 5 & 10.77 $\pm$ 0.19 & 6 \\ 
NGC 2549 & 7.15 (+0.06, -1.15) & 2 & 9.94 $\pm$ 0.15 & 3 & 10.01 $\pm$ 0.30 & 7 \\ 
NGC 2748 & 7.54 (+0.15, -0.23) & 1 & --- & 5 & 10.09 $\pm$ 0.22 & 6 \\ 
NGC 2778 & 7.18 (+0.20, -0.48) & 2 & 9.40 (+0.24, -0.28) & 2 & 10.66 $\pm$ 0.30 & 8 \\ 
NGC 2787 & 7.60 $\pm$ 0.06 & 4 & 9.13 $\pm$ 0.26 & 4 & 9.99 $\pm$ 0.12 & 4 \\ 
NGC 2960 & 7.06 $\pm$ 0.03 & 1 & 10.44 $\pm$ 0.36 & 5 & 10.86 $\pm$ 0.34 & 6 \\ 
NGC 2974 & 8.23 $\pm$ 0.05 & 1 & 10.23 $\pm$ 0.13 & 5 & 10.73 $\pm$ 0.12 & 6 \\ 
NGC 3031 & 7.83 (+0.11, -0.07) & 1 & 10.16 $\pm$ 0.11 & 5 & 10.65 $\pm$ 0.08 & 6 \\ 
NGC 3079 & 6.38 (+0.08, -0.10) & 1 & 9.92 $\pm$ 0.25 & 5 & 10.68 $\pm$ 0.18 & 6 \\ 
NGC 3091 & 9.56 (+0.01, -0.02) & 2 & 11.48 (+0.04, -0.08) & 2 & 11.29 $\pm$ 0.30 & 7 \\ 
NGC 3115 & 8.94 (+0.33, -0.16) & 2 & 10.19 $\pm$ 0.15 & 3 & 10.64 $\pm$ 0.30 & 7 \\ 
NGC 3227 & 7.86 (+0.17, -0.25) & 1 & 10.04 $\pm$ 0.17 & 5 & 10.80 $\pm$ 0.14 & 6 \\ 
NGC 3245 & 8.30 (+0.10, -0.12) & 2 & 10.44 $\pm$ 0.15 & 3 & 10.50 $\pm$ 0.30 & 7 \\ 
NGC 3368 & 6.89 (+0.08, -0.10) & 1 & 9.81 $\pm$ 0.10 & 5 & 10.69 $\pm$ 0.09 & 6 \\ 
NGC 3377 & 7.88 (+0.02, -0.04) & 2 & 9.96 $\pm$ 0.15 & 3 & 10.14 $\pm$ 0.30 & 7 \\ 
NGC 3379 (M105) & 8.60 (+0.10, -0.12) & 2 & 10.67 $\pm$ 0.15 & 3 & 10.59 $\pm$ 0.30 & 7 \\ 
NGC 3384 & 7.23 (+0.02, -0.05) & 2 & 10.20 $\pm$ 0.15 & 3 & 10.46 $\pm$ 0.30 & 7 \\ 
NGC 3393 & 7.49 (+0.05, -0.06) & 1 & 10.23 $\pm$ 0.12 & 5 & 11.00 $\pm$ 0.10 & 6 \\ 
NGC 3414 & 8.38 (+0.05, -0.06) & 2 & 10.47 $\pm$ 0.15 & 3 & 10.95 $\pm$ 0.30 & 8 \\ 
NGC 3489 & 6.76 $\pm$ 0.06 & 2 & 9.62 (+0.23, -0.26) & 2 & 10.21 $\pm$ 0.30 & 7 \\ 
NGC 3585 & 8.49 (+0.16, -0.09) & 2 & 10.95 $\pm$ 0.15 & 3 & 10.96 $\pm$ 0.30 & 7 \\ 
NGC 3607 & 8.11 (+0.14, -0.21) & 2 & 10.90 $\pm$ 0.15 & 3 & 10.93 $\pm$ 0.30 & 7 \\ 
NGC 3608 & 8.30 (+0.19, -0.15) & 2 & 10.61 $\pm$ 0.15 & 3 & 10.69 $\pm$ 0.30 & 7 \\ 
NGC 3627 & 6.95 $\pm$ 0.05 & 1 & 9.74 $\pm$ 0.20 & 5 & 10.78 $\pm$ 0.10 & 6 \\ 
NGC 3665 & 8.76 $\pm$ 0.10 & 4 & 11.03 $\pm$ 0.26 & 4 & 11.28 $\pm$ 0.12 & 4 \\ 
NGC 3842 & 9.99 (+0.12, -0.14) & 2 & 11.79 (+0.05, -0.07) & 2 & 11.44 $\pm$ 0.30 & 7 \\ 
NGC 3923 & 9.45 $\pm$ 0.13 & 4 & 11.4 $\pm$ 0.15 & 4 & 11.40 $\pm$ 0.12 & 4 \\ 
NGC 3998 & 8.91 (+0.10, -0.12) & 2 & 10.66 $\pm$ 0.15 & 3 & 10.41 $\pm$ 0.30 & 7 \\ 
NGC 4026 & 8.26 $\pm$ 0.11 & 4 & 10.11 $\pm$ 0.33 & 4 & 10.36 $\pm$ 0.12 & 4 \\ 
NGC 4151 & 7.68 (+0.15, -0.60) & 1 & 10.27 $\pm$ 0.15 & 5 & 10.62 $\pm$ 0.14 & 6 \\ 
NGC 4258 & 7.60 $\pm$ 0.01 & 1 & 10.05 $\pm$ 0.18 & 5 & 10.72 $\pm$ 0.09 & 6 \\ 
NGC 4261 & 8.70 (+0.08, -0.10) & 2 & 11.19 $\pm$ 0.15 & 3 & 11.33 $\pm$ 0.30 & 7 \\ 
NGC 4291 & 8.52 (+0.10, -0.62) & 2 & 10.55 $\pm$ 0.15 & 3 & 10.52 $\pm$ 0.30 & 7 \\ 
NGC 4303 & 6.58 (+0.07, -0.26) & 1 & 9.42 $\pm$ 0.10 & 5 & 10.48 $\pm$ 0.09 & 6 \\ 
NGC 4339 & 7.63 $\pm$ 0.33 & 4 & 9.67 $\pm$ 0.26 & 4 & 10.17 $\pm$ 0.12 & 4 \\ 
NGC 4342 & 8.65 $\pm$ 0.18 & 4 & 9.94 $\pm$ 0.25 & 4 & 10.26 $\pm$ 0.12 & 4 \\ 
NGC 4350 & 8.86 $\pm$ 0.41 & 4 & 10.28 $\pm$ 0.26 & 4 & 10.55 $\pm$ 0.12 & 4 \\ 
NGC 4371 & 6.84 $\pm$ 0.08 & 4 & 9.89 $\pm$ 0.26 & 4 & 10.60 $\pm$ 0.12 & 4 \\ 
NGC 4374 (M84) & 8.95 $\pm$ 0.04 & 2 & 11.45 (+0.23, -0.27) & 2 & 11.29 $\pm$ 0.30 & 7 \\ 
NGC 4388 & 6.90 (+0.04, -0.05) & 1 & 10.07 $\pm$ 0.22 & 5 & 10.44 $\pm$ 0.22 & 6 \\ 
NGC 4395 & 5.64 (+0.22, -0.12) & 1 & --- & 5 & 9.45 $\pm$ 0.08 & 6 \\ 
NGC 4429 & 8.18 $\pm$ 0.09 & 4 & 10.46 $\pm$ 0.26 & 4 & 10.90 $\pm$ 0.12 & 4 \\ 
NGC 4434 & 7.84 $\pm$ 0.17 & 4 & 9.91 $\pm$ 0.26 & 4 & 10.18 $\pm$ 0.12 & 4 \\ 
NGC 4459 & 7.83 (+0.08, -0.09) & 2 & 10.36 $\pm$ 0.15 & 3 & 10.56 $\pm$ 0.30 & 7 \\ 
NGC 4472 (M49) & 9.40 (+0.05, -0.02) & 2 & 11.59 (+0.04, -0.07) & 2 & 11.51 $\pm$ 0.30 & 7 \\ 
NGC 4473 & 8.08 (+0.12, -0.60) & 2 & 10.64 $\pm$ 0.15 & 3 & 10.55 $\pm$ 0.30 & 7 \\ 
NGC 4486 (M87) & 9.76 $\pm$ 0.03 & 2 & 11.28 $\pm$ 0.15 & 3 & 11.38 $\pm$ 0.30 & 7 \\ 
NGC 4486A & 7.11 (+0.12, -0.34) & 3 & 10.06 $\pm$ 0.15 & 3 & 9.71 $\pm$ 0.30 & 7 \\ 
NGC 4486B & 8.76 $\pm$ 0.24 & 4 & 9.46 $\pm$ 0.33 & 4 & 9.46 $\pm$ 0.12 & 4 \\ 
NGC 4501 & 7.13 $\pm$ 0.08 & 1 & 10.11 $\pm$ 0.16 & 5 & 10.67 $\pm$ 0.08 & 6 \\ 
NGC 4526 & 8.67 $\pm$ 0.04 & 4 & 10.70 $\pm$ 0.26 & 4 & 11.04 $\pm$ 0.12 & 4 \\ 
NGC 4552 (M89) & 8.67 $\pm$ 0.05 & 4 & 10.88 $\pm$ 0.25 & 4 & 10.95 $\pm$ 0.12 & 4 \\ 
NGC 4564 & 7.78 (+0.02, -0.07) & 2 & 10.10 $\pm$ 0.15 & 3 & 10.23 $\pm$ 0.30 & 7 \\ 
NGC 4578 & 7.28 $\pm$ 0.35 & 4 & 9.77 $\pm$ 0.26 & 4 & 10.23 $\pm$ 0.12 & 4 \\ 
NGC 4594 & 8.81 $\pm$ 0.08 & 1 & 10.81 $\pm$ 0.20 & 5 & 11.03 $\pm$ 0.14 & 6 \\ 
NGC 4596 & 7.90 (+0.17, -0.28) & 2 & 10.19 $\pm$ 0.15 & 3 & 10.48 $\pm$ 0.30 & 7 \\ 
NGC 4621 (M59) & 8.59 (+0.04, -0.05) & 2 & 10.53 $\pm$ 0.15 & 3 & 10.97 $\pm$ 0.30 & 8 \\ 
NGC 4649 & 9.67 $\pm$ 0.10 & 4 & 11.44 $\pm$ 0.12 & 4 & 11.44 $\pm$ 0.12 & 4 \\ 
NGC 4697 & 8.26 (+0.05, -0.02) & 2 & 10.28 $\pm$ 0.15 & 3 & 10.63 $\pm$ 0.30 & 7 \\ 
NGC 4699 & 8.34 $\pm$ 0.05 & 1 & 11.12 $\pm$ 0.26 & 5 & 11.29 $\pm$ 0.23 & 6 \\ 
NGC 4736 & 6.78 (+0.09, -0.11) & 1 & 9.89 $\pm$ 0.09 & 5 & 10.37 $\pm$ 0.08 & 6 \\ 
NGC 4742 & 7.15 $\pm$ 0.18 & 4 & 9.87 $\pm$ 0.26 & 4 & 10.15 $\pm$ 0.12 & 4 \\ 
NGC 4751 & 9.15 $\pm$ 0.05 & 4 & 10.49 $\pm$ 0.26 & 4 & 10.72 $\pm$ 0.12 & 4 \\ 
NGC 4762 & 7.36 $\pm$ 0.15 & 4 & 9.97 $\pm$ 0.28 & 4 & 11.06 $\pm$ 0.12 & 4 \\ 
NGC 4826 & 6.07 (+0.10, -0.12) & 1 & 9.55 $\pm$ 0.22 & 5 & 10.41 $\pm$ 0.21 & 6 \\ 
NGC 4889 & 10.32 (+0.25, -0.62) & 2 & 11.96 (+0.05, -0.07) & 2 & 11.81 $\pm$ 0.30 & 7 \\ 
NGC 4945 & 6.15 $\pm$ 0.30 & 1 & 9.39 $\pm$ 0.19 & 5 & 10.52 $\pm$ 0.09 & 6 \\ 
NGC 5018 & 8.02 $\pm$ 0.09 & 4 & 10.98 $\pm$ 0.27 & 4 & 11.35 $\pm$ 0.12 & 4 \\ 
NGC 5055 & 8.94 (+0.09, -0.11) & 1 & 10.49 $\pm$ 0.11 & 5 & 10.81 $\pm$ 0.10 & 6 \\ 
NGC 5077 & 8.87 (+0.21, -0.23) & 2 & 11.03 $\pm$ 0.15 & 3 & 10.99 $\pm$ 0.30 & 7 \\ 
NGC 5128 & 7.65 (+0.14, -0.11) & 2 & 10.09 $\pm$ 0.15 & 3 & 10.73 $\pm$ 0.30 & 7 \\ 
NGC 5252 & 9.00 $\pm$ 0.40 & 4 & 10.85 $\pm$ 0.26 & 4 & 11.38 $\pm$ 0.12 & 4 \\ 
NGC 5328 & 9.67 $\pm$ 0.15 & 4 & 11.49 $\pm$ 0.12 & 4 & 11.49 $\pm$ 0.12 & 4 \\ 
NGC 5419 & 9.86 $\pm$ 0.14 & 4 & 11.44 $\pm$ 0.12 & 4 & 11.44 $\pm$ 0.12 & 4 \\ 
NGC 5495 & 7.04 (+0.08, -0.09) & 1 & 10.54 $\pm$ 0.12 & 5 & 11.31 $\pm$ 0.10 & 6 \\ 
NGC 5516 & 9.52 $\pm$ 0.06 & 4 & 11.44 $\pm$ 0.12 & 4 & 11.44 $\pm$ 0.12 & 4 \\ 
NGC 5576 & 8.20 (+0.07, -0.12) & 2 & 10.28 $\pm$ 0.15 & 3 & 10.64 $\pm$ 0.30 & 7 \\ 
NGC 5765b & 7.72 $\pm$ 0.03 & 1 & 10.04 $\pm$ 0.13 & 5 & 11.11 $\pm$ 0.12 & 6 \\ 
NGC 5813 & 8.83 $\pm$ 0.06 & 4 & 10.86 $\pm$ 0.26 & 4 & 11.23 $\pm$ 0.12 & 4 \\ 
NGC 5845 & 8.41 $\pm$ 0.22 & 4 & 10.12 $\pm$ 0.26 & 4 & 10.32 $\pm$ 0.12 & 4 \\ 
NGC 5846 & 9.04 $\pm$ 0.04 & 2 & 11.10 $\pm$ 0.15 & 3 & 11.40 $\pm$ 0.30 & 8 \\ 
NGC 6086 & 9.57 $\pm$ 0.16 & 4 & 11.52 $\pm$ 0.26 & 4 & 11.52 $\pm$ 0.12 & 4 \\ 
NGC 6251 & 8.77 (+0.15, -0.22) & 2 & 11.66 (+0.04, -0.07) & 2 & 11.55 $\pm$ 0.30 & 7 \\ 
NGC 6264 & 7.51 $\pm$ 0.02 & 1 & 10.01 $\pm$ 0.15 & 5 & 11.06 $\pm$ 0.14 & 6 \\ 
NGC 6323 & 7.02 $\pm$ 0.02 & 1 & 9.86 $\pm$ 0.31 & 5 & 11.04 $\pm$ 0.28 & 6 \\ 
NGC 6861 & 9.30 $\pm$ 0.08 & 4 & 10.94 $\pm$ 0.29 & 4 & 11.02 $\pm$ 0.12 & 4 \\ 
NGC 6926 & 7.74 (+0.26, -0.74) & 1 & --- & 5 & 11.31 $\pm$ 0.08 & 6 \\ 
NGC 7052 & 8.57 $\pm$ 0.23 & 4 & 11.46 $\pm$ 0.12 & 4 & 11.46 $\pm$ 0.12 & 4 \\ 
NGC 7332 & 7.11 $\pm$ 0.20 & 4 & 10.22 $\pm$ 0.34 & 4 & 10.84 $\pm$ 0.12 & 4 \\ 
NGC 7457 & 7.00 $\pm$ 0.30 & 4 & 9.40 $\pm$ 0.26 & 4 & 10.19 $\pm$ 0.12 & 4 \\ 
NGC 7582 & 7.67 (+0.09, -0.08) & 1 & 10.15 $\pm$ 0.20 & 5 & 10.77 $\pm$ 0.11 & 6 \\ 
NGC 7619 & 9.40 (+0.12, -0.06) & 2 & 11.52 (+0.23, -0.26) & 2 & 11.34 $\pm$ 0.30 & 7 \\ 
NGC 7768 & 9.11 (+0.14, -0.16) & 2 & 11.76 (+0.20, -0.26) & 2 & 11.40 $\pm$ 0.30 & 7 \\ 
UGC 3789 & 7.06 (+0.02, -0.03) & 1 & 10.18 $\pm$ 0.14 & 5 & 10.74 $\pm$ 0.13 & 6 \\ 
UGC 6093 & 7.45 $\pm$ 0.04 & 1 & 10.35 $\pm$ 0.14 & 5 & 11.26 $\pm$ 0.11 & 6 \\ 
\enddata
\tablecomments{Columns: Same as table \ref{tab:BH_relations_reverb} but with $M_{\rm{BH}}$ estimates from dynamical measurements.\\
References: (1) \citep{Davis2017}. (2) \citep{Savorgnan2016}. (3) \citep{Hu2009}. (4) \citep{Sahu2019}. (5) \citep{Davis2019}. (6) \citep{Davis2018}. (7) \citep{Reines2015}. (8) \citep{Dabringhausen2016}.}
\end{deluxetable*}

\pagebreak

\bibliography{Paper.bib}
\bibliographystyle{aasjournal}

\end{document}